\documentclass[utf8]{FrontiersinHarvard} 

\usepackage{url,hyperref,lineno,microtype}
\usepackage[onehalfspacing]{setspace}

\usepackage{enumitem}
\setlist{labelsep=0.5em,leftmargin=*}
\usepackage{fix-cm} 
\usepackage{xcolor}

\renewcommand{\cite}{\citep}

\def\keyFont{\fontsize{8}{11}\helveticabold }
\def\firstAuthorLast{Harris {et~al.}}

\def\Authors{Megan S. Harris\,$^{1,\dagger,*}$, Mohammad Mahdi Naderi\,$^{1}$, Ehsanoddin Ghorbanichemazkati\,$^{2}$, John C. Little\,$^{1}$, and Amro M. Farid\,$^{2}$}
\def\Address{$^{1}$Department of Civil and Environmental Engineering, Virginia Tech, Blacksburg, VA, USA \\
$^{2}$School of Systems and Enterprises, Stevens Institute of Technology, Hoboken, NJ, USA \\
$^{\dagger}$Present address: Department of Civil, Construction and Environmental Engineering, North Carolina State University, Raleigh, NC, USA}
\def\corrAuthor{Megan S. Harris}
\def\corrEmail{megan\_harris@ncsu.edu}
\usepackage{float}
\allowdisplaybreaks

\usepackage{amsthm}
\usepackage{thmtools}
\usepackage{xurl}
\declaretheoremstyle[%
  headfont=\bfseries,%
  headpunct={:},%
  notefont=\normalfont\bfseries,%
  notebraces={--~}{},%
  qed=$\blacksquare$,%
]{definitionstyle}

\declaretheorem[style=definitionstyle,name=Definition]{defn}

\theoremstyle{definition}

\theoremstyle{remark}

\usepackage{nicematrix}
\NiceMatrixOptions{
code-for-first-row = \color{blue} ,
code-for-last-row = \color{blue} ,
code-for-first-col = \color{blue} ,
code-for-last-col = \color{blue}
}
\newcommand{\bibAnnote}[2]{}\newcommand{\bibAnnoteFile}[1]{}

\begin{document}
\onecolumn
\firstpage{1}

\title[Hetero-functional Graph Theory for Hydrology]{An Introduction to Model-Based Systems Engineering and Hetero-functional Graph Theory for Hydrological Systems of Systems}

\author[\firstAuthorLast ]{\Authors}
\address{}
\correspondance{}
\extraAuth{}

\maketitle

\begin{abstract}

\section{}
This introductory overview describes a unified modeling framework for representing interconnected environmental systems using Model-Based Systems Engineering and Hetero-functional Graph Theory. Existing environmental models encode rich structural and process-level information, yet much of this knowledge remains implicit within domain-specific formulations. This paper demonstrates how such information can be formalized into a consistent ontology that explicitly represents system structure, function, and interconnections. Using illustrative hydrological examples, the tutorial illustrates how conventional process models can be translated into a scalable system-of-systems representation. Agricultural nitrogen applied to land is represented explicitly as the point at which human decisions enter the hydrological system, illustrating how models of human systems can be coupled to hydrological models within the same representation. These examples highlight how embedded relationships in existing models can be made explicit and computationally tractable. Using simplified examples, the overview provides a foundation for integrating heterogeneous models across environmental and engineered domains. More broadly, this work offers a step-by-step introduction to a generalizable modeling framework for complex systems.

\tiny
 \keyFont{ \section{Keywords:} environmental systems modeling, modeling framework, systems of systems (SoS), Model-Based Systems Engineering (MBSE), systems modeling language (SysML), Hetero-functional Graph Theory (HFGT)}
\end{abstract}

\section{Introduction}
Environmental and engineering challenges rarely exist within a single discipline or infrastructure system. Water resource management, flood resilience, agricultural production, and energy supply arise from interactions among hydrological, ecological, engineered, social, and regulatory systems~\cite{Biermann:2022:00,reed:2022:01}. Understanding these interactions therefore requires modeling approaches capable of representing not only individual systems, but also the larger \emph{systems of systems} in which they are embedded. As these systems become increasingly interconnected, there is a growing need for methodologies that can consistently represent multiple interacting systems while preserving the unique characteristics of each domain~\cite{Biermann:2022:00,bodin:2017:00,reed:2008:00,zare:2024:00,Little:2023:ISC-J50}.

\vspace{0.5em} \begin{defn}[System of Systems]\label{defn:cbw-SoS} 1) System of interest whose system elements are themselves systems; typically, these entail large-scale interdisciplinary problems with multiple, heterogeneous, distributed systems~\cite{SE-Handbook-Working-Group:2015:00}. 2) Set of systems or system elements that interact to provide a unique capability that none of the constituent systems can accomplish on its own~\cite{ISO-IEC-IEEE:2019:00}. \end{defn} \vspace{0.5em}

Decades of progress in environmental modeling have produced sophisticated process models for representing hydrological, geochemical, ecological, atmospheric, and other Earth systems~\cite{bennett:2013:00,keller:2023:00}. These models provide detailed, theory-driven representations of individual systems and encode knowledge of system structure, process connectivity, and causal relationships beyond the mathematical equations used to simulate them~\cite{bennett:2013:00,keller:2023:00,nativi:2013:00,visser:2019:00}. The diversity of these modeling approaches reflects the richness of environmental science, as each has been developed to answer domain-specific questions using its own assumptions, variables, mathematical formulations, and data representations~\cite{Little:2019:00,Guizzardi:2005:00}. Consequently, existing process models are both data- and theory-rich, providing valuable insights within their respective disciplines.

This specialization, however, presents a challenge when environmental problems span multiple interacting systems. Much of the structural knowledge embedded within existing process models remains implicit within parameterizations, calibration choices, model couplings, or software implementations rather than being represented explicitly in a form that can be readily shared or reused across disciplines~\cite{nativi:2013:00,keller:2023:00,hughes:2022:00,raming:2025:00,salas:2020:00}. As a result, valuable information encoded in one system model cannot easily be integrated with another, especially when the underlying systems interact in the real world~\cite{folke:2006:00,Farid:2022:ISC-C79,Bi:2022:00}. Integrated approaches such as system dynamics address cross-system feedbacks directly, but typically do so by aggregating processes into a small number of stocks and flows, sacrificing the spatial resolution and process detail of the underlying domain models~\cite{Naderi:2026:ISC-JR08}. Developing modeling methodologies that preserve domain-specific knowledge while providing a common representation across disciplines is therefore an important step toward integrated analysis of environmental systems of systems~\cite{meirer:2025:00,Little:2026:ISC-JR10}.

One barrier to developing such integrated models is the lack of a shared conceptual language for describing systems. Models from different disciplines often define similar concepts using different terminology or represent comparable processes using incompatible abstractions. In this paper, this common conceptual language is referred to as an \emph{ontology}.

\vspace{0.5em} \begin{defn}[Ontology~\cite{gruber:1995:00,Guizzardi:2005:00}]\label{defn:ontology}
An explicit specification of a conceptualization that defines the concepts,
relationships, and constraints used to represent knowledge within a particular
domain.
\end{defn}

\subsection{Paper Objective}
This methods paper introduces \emph{Hetero-functional Graph Theory} (HFGT) as a methodology for constructing ontology-based models of environmental systems of systems. Rather than replacing existing process models, the methodology provides a common representational framework that makes their underlying structure explicit while preserving the scientific knowledge embedded within them. The approach combines Model-Based Systems Engineering (MBSE) with HFGT to formally represent system elements, processes, and their interactions within a consistent modeling framework~\cite{Little:2019:00,Farid:2022:ISC-C79}. Once represented within this common ontology, heterogeneous system models can be analyzed using a unified mathematical framework while remaining faithful to their domain-specific formulations.

\vspace{0.5em} \begin{defn}[Hetero-functional Graph Theory (HFGT)~\cite{Farid:2022:ISC-J49,Farid:2025:ISC-JR06}]\label{defn:hfgt}
A modeling methodology that fuses Model-Based Systems Engineering (MBSE) and network science by preserving the heterogeneous conceptual and ontological constructs of engineering systems within graph-based mathematical representations for analysis and simulation
\end{defn} \vspace{0.5em}

The objective of this tutorial is to introduce the methodology rather than to provide a comprehensive treatment of HFGT, which is given by Farid et al.~\cite{Farid:2025:ISC-JR06}. Illustrative hydrological examples provide an intuitive setting for introducing and learning the fundamental concepts, workflow, and mathematical representations. Although the demonstration focuses on hydrological systems, the methodology is equally applicable to a broad range of environmental, engineered, and socio-technical systems. Within this example, nitrogen applied to agricultural land enters the hydrological system through an explicit capability of each land segment; this capability is the interface through which models of human systems, such as agricultural production or land management decisions, can later be coupled to the hydrological model. Throughout the paper, references to applications in domains such as energy systems, water infrastructure, transportation, manufacturing, and healthcare are provided to illustrate the generality of the approach and guide readers toward more advanced applications.

\subsection{Paper Outline}
The remainder of this paper follows the Frontiers Methods structure. Section~\ref{sec:materials} lists the software and data required to apply the methodology. Section~\ref{sec:methods} presents the method: it summarizes the step-by-step workflow, introduces the pedagogical hydrological system that serves as the running example, presents the HFGT meta-architecture and its foundational concepts, and develops the mathematical formulation of HFGT together with its hydrological counterpart. Section~\ref{sec:instantiation} presents the results of instantiating the methodology for hydrological systems of progressively increasing complexity. Section~\ref{sec:discussion} discusses the implications, limitations, and recommended modeling practices of the methodology, and Section~\ref{sec:conclusion} concludes the paper and highlights directions for future work.

\section{Materials and Equipment} \label{sec:materials}
The methodology requires only software and a standard personal computer; no specialized hardware or laboratory equipment is needed. The following tools were used:
\begin{itemize}[noitemsep]
    \item \textbf{SysML modeling:} Block Definition Diagrams and Activity Diagrams describing the reference architecture (Figures~\ref{fig:LFESMetaArchitecturebdd}--\ref{fig:hydro-act}) were developed in SysML~v1 using Magic Systems of Systems Architect 2024x (No Magic, Inc., a Dassault Syst\`emes company).
    \item \textbf{HFGT toolbox:} The HFGT toolbox~\cite{Farid:2025:ISC-BKR01} reads an XML description of an instantiated architecture and generates the hetero-functional incidence tensors and associated system metadata in HDF5 format. Version 4.0.8 of the toolbox, implemented in Python and released under the MIT license, was used for this paper and is archived in the accompanying repository; it was run with Python 3.11, SciPy 1.14.1, and GNU Octave, which the toolbox uses to write HDF5 files. Newer versions of the toolbox are available from the authors upon request.
    \item \textbf{Simulation environment:} The toolbox outputs were loaded into Julia (version 1.12.4) using HDF5.jl, and the instantiated HFNMCF model was assembled with Julia's SparseArrays and LinearAlgebra standard libraries. Because the objective function is zero and every capability firing is fixed either by an exogenous input or by a device model (Section~\ref{sec:hfgtmath}), no optimization solver is required: the model is advanced by explicit time-stepping of the Engineering System Net state transition function with $\Delta T = 6$~h over 52 weeks. For applications with a nonzero objective function or flows left free as decision variables, the same constraints define an optimization program, which is nonlinear in general because the mixing constraints (Eqs.~\ref{Eq:simplified-mixing-land}--\ref{Eq:simplified-mixing-point}) are bilinear; such problems can be posed in an algebraic modeling layer such as JuMP~\cite{Lubin:2023:00} and solved with a nonlinear solver such as Ipopt or with Gurobi's nonconvex quadratic capability. Exact Julia package versions are recorded in the repository. The continuous-time reference solution for Illustrative Example 1 was computed with DifferentialEquations.jl using the Tsit5 integrator. Figures were produced with StatsPlots.jl.
    \item \textbf{Input data:} All parameter values and exogenous inputs were synthetically generated for illustrative purposes, scaled to published precipitation, evapotranspiration, and nitrogen loading data (Section~\ref{sec:parameters}).
\end{itemize}
All XML input files, toolbox outputs, Julia scripts, and figure-generation scripts are available in the repository listed in the Data Availability Statement.

\section{Methods} \label{sec:methods}

\subsection{Overview of the Modeling Workflow} \label{sec:workflow}
The MBSE--HFGT methodology proceeds through the following steps, each of which is developed in the remainder of this section and demonstrated in Section~\ref{sec:instantiation}:
\begin{enumerate}[noitemsep]
    \item Define the system of interest (Section~\ref{sec:runningexample}).
    \item Specialize the HFGT meta-architecture into a domain-specific reference architecture by defining resources, buffers, processes, and capabilities as SysML Block Definition and Activity Diagrams (Section~\ref{sec:hfgtterms}).
    \item Instantiate the reference architecture for a specific system and encode it in an XML file (Section~\ref{sec:instantiation}).
    \item Process the XML file with the HFGT toolbox to generate the hetero-functional incidence tensors and system metadata (Section~\ref{sec:hfgtmath}).
    \item Assemble the HFNMCF problem, adding the domain-specific device models, boundary conditions, and initial conditions (Section~\ref{sec:hfgtmath}).
    \item Simulate the system and verify the results against an independent formulation where one is available (Section~\ref{sec:instantiation}).
\end{enumerate}

Steps~1--3 involve modeling, with the effort required depending primarily on the modeler's familiarity with SysML and HFGT and the availability of a domain-specific reference architecture. Developing a new reference architecture (Step~2) is the most substantial task, but subsequent updates require less effort as new resources, processes, or operands are identified. Once the architecture is established, instantiating a new system (Step~3) requires only specifying its elements and connections. Steps~4--6 are computational and, for the examples in this paper, each completed in seconds to minutes on a standard laptop. Because the XML file generated in Step~3 and the toolbox output generated in Step~4 are stored, the workflow can be paused after either step and resumed later. Reference architectures and instantiated systems can also be reused without repeating earlier steps.

\subsection{Running Example: A Pedagogical Hydrological System} \label{sec:runningexample}

Throughout this paper, a simple hydrological system serves as a running example for illustrating the concepts and methodology of HFGT. The system is intentionally simplified so that each modeling concept can be introduced incrementally without the complexity of a realistic watershed. As new HFGT concepts are presented, they are immediately clarified using this example, demonstrating how a physical system is progressively transformed into an executable mathematical model.

\begin{figure}[htbp]
    \centering
    \includegraphics[width=0.8\linewidth]{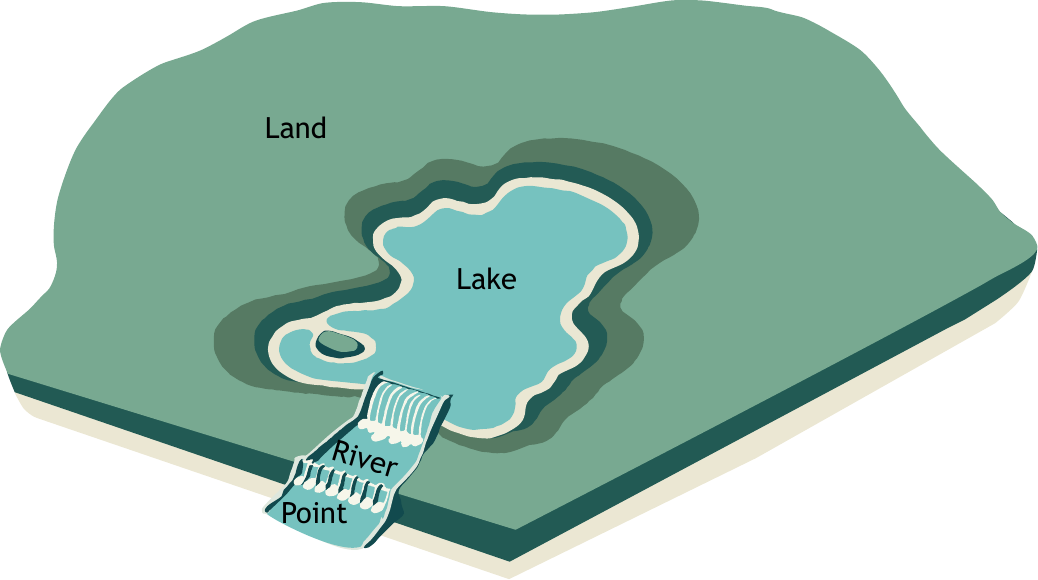}
    \caption{Diagram of the pedagogical hydrologic system.}
    \label{fig:genericlakeland}
\end{figure}

Figure~\ref{fig:genericlakeland} illustrates the pedagogical hydrological system. The system consists of land segments, lakes, rivers, and river points that exchange water and nitrogen. Water enters land segments through precipitation, while nitrogen is introduced through fertilizer application. Runoff transports water and dissolved nitrogen from land segments to downstream lakes, where they are stored and assumed to mix completely. Rivers then transport water and nitrogen between lakes and river points, which represent locations where flows merge, split, or leave the modeled watershed. Together these elements provide a simplified watershed representation that captures the essential processes of storage, transport, and mixing needed to illustrate the HFGT methodology.

Before introducing the HFGT representation, it is helpful to consider how this system would traditionally be modeled using mass-balance equations. Under the assumption of complete mixing within each storage element, the system can be viewed as a network of Continuous Stirred Tank Reactors (CSTRs)~\cite{kumar:2012:01}. A traditional environmental engineering formulation therefore consists of the following coupled differential equations:
\begin{subequations}
\begin{align}
\frac{dV_{\text{land}}}{dt} &= \dot{V}_{\text{rain,land}} - \dot{V}_{\text{runoff}} \label{Eq:dif-MB-water-land} \\
\frac{dV_{\text{lake}}}{dt} &= \dot{V}_{\text{rain,lake}} + \dot{V}_{\text{runoff}} - \dot{V}_{\text{riverflow}} \label{Eq:dif-MB-water} \\
\frac{dV_{\text{point}}}{dt} &= \dot{V}_{\text{riverflow}} \label{Eq:dif-MB-water-point} \\
\frac{dm_{land}}{dt} &= \dot{m}_{\text{fertilizer}}-\dot{m}_{\text{runoff}} \label{Eq:dif-MB-nitrogen-land}\\
\frac{dm_{lake}}{dt} &= {\dot{m}_{\text{runoff}}}-\dot{m}_{\text{riverflow}} \label{Eq:dif-MB-nitrogen}\\
\frac{dm_{point}}{dt} &= \dot{m}_{\text{riverflow}} \label{Eq:dif-MB-nitrogen-point}\\
\dot{V}_{\text{runoff}} &= \frac{\rho g}{R}\left(\frac{V_{\text{land}}}{A_{\text{land}}} - \frac{V_{\text{lake}}}{A_{\text{lake}}} + z_{\text{land}} -z_{\text{lake}}\right) \label{Eq:dif-resistance-land} \\
\dot{V}_{\text{riverflow}} &= \frac{\rho g}{R}\left(\frac{V_{\text{lake}}}{A_{\text{lake}}} - \frac{V_{\text{point}}}{A_{\text{point}}} + z_{\text{lake}} -z_{\text{point}}\right) \label{Eq:dif-resistance} \\
\dot{m}_{\text{runoff}} &= \frac{m_{land}}{V_{\text{land}}}\dot{V}_{\text{runoff}} \label{Eq:dif-mixing-land} \\
\dot{m}_{\text{riverflow}} &= \frac{m_{lake}}{V_{\text{lake}}}\dot{V}_{\text{riverflow}} \label{Eq:dif-mixing}
\end{align}
\end{subequations}
where:
\begin{itemize}
\item Equations~\ref{Eq:dif-MB-water-land}--\ref{Eq:dif-MB-nitrogen-point} represent the \textbf{mass balance} of water and nitrogen in the land, lake, and outlet point.
\item Equations \ref{Eq:dif-resistance-land} and \ref{Eq:dif-resistance} define the \textbf{flow of water downstream} modulated by a resistance term based on the hydraulic pressure differential between upstream and downstream elements.
\item Equations \ref{Eq:dif-mixing-land} and \ref{Eq:dif-mixing} describe the \textbf{constituent transport} of nitrogen from the land and the lake under the assumption of complete mixing, such that the nitrogen concentration within each element equals the concentration out.
\end{itemize}

Although this system is small enough to formulate explicitly using differential equations, it contains many of the fundamental characteristics shared by environmental and infrastructure systems, including material storage, transport, mixing, and transformation. As a result, the HFGT concepts introduced using this pedagogical example generalize naturally to larger watersheds and, more broadly, to other classes of engineering systems. The objective of this paper is not to replace the governing equations of hydrology, but rather to demonstrate how an equivalent mathematical model can be derived systematically from an explicit representation of system structure and function.

\subsection{Hetero-functional Graph Theory Meta-Architecture} \label{sec:hfgtterms}

The previous section introduced the physical hydrological system that serves as the running example throughout this methods paper. HFGT provides a domain-independent modeling methodology for representing such systems using a common ontology. Rather than beginning with hydrological concepts such as lakes, rivers, or runoff, HFGT first defines a generic modeling language that can describe any engineering system. This language, referred to as the HFGT \textit{meta-architecture}, specifies the fundamental concepts used to represent system structure and behavior. 

\vspace{0.5em} \begin{defn}[Meta-Architecture~\cite{ISO-IEC-IEEE:2019:00,gruber:1995:00,mohammad:2008:00}]
A domain-independent ontology that defines the concepts, relationships, and constraints used to describe a class of engineering systems.
\end{defn} \vspace{0.5em}

A domain-specific \emph{reference architecture} is developed by specializing the HFGT meta-architecture for a particular class of engineering systems. For instance, this paper elaborates on the domain of hydrology. Individual engineering systems are then modeled by instantiating the reference architecture with specific system elements and their interconnections to create an \emph{instantiated architecture}. Individual hydrological systems are instantiated in Section \ref{sec:instantiation} of this manuscript.

\vspace{0.5em} \begin{defn}[Reference Architecture~\cite{cloutier:2010:00,ISO-IEC-IEEE:2019:00}]
A standardized architecture for a particular class of systems that captures their common concepts, relationships, and architectural patterns and serves as the basis for constructing specific system architectures.
\end{defn} 
\vspace{0.5em}

\begin{defn}[Instantiated Architecture~\cite{Thompson:2022:00}]
A specific realization of a reference architecture in which the architectural
elements are assigned to real world system components and their interconnections.
\end{defn} \vspace{0.5em}

\begin{figure}[htbp]
    \centering
    \includegraphics[width=0.95\linewidth]{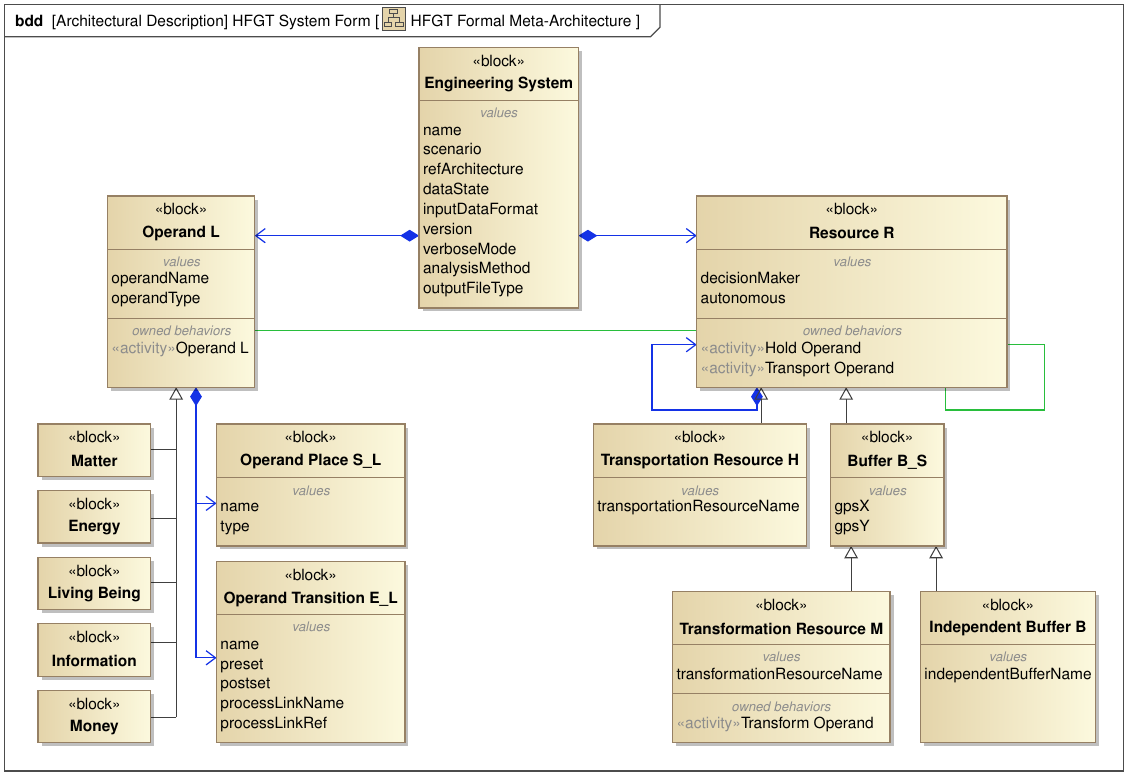}
    \caption{A SysML Block Definition Diagram of the System Form of the System Meta-Architecture. Adapted from \protect{\cite{Schoonenberg:2019:00}}.}
    \label{fig:LFESMetaArchitecturebdd}
\end{figure}

\begin{figure}[htbp]
    \centering
    \includegraphics[width=0.95\linewidth]{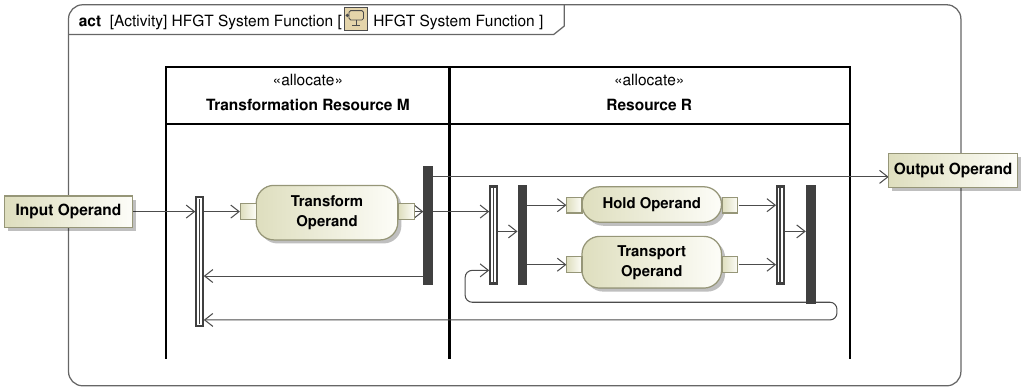}
    \caption{A SysML Activity Diagram of the System Function of the System Meta-Architecture. Adapted from \protect{\cite{Schoonenberg:2019:00}}.}
    \label{fig:LFESMetaArchitectureact}
\end{figure}

\begin{figure}[htbp]
    \centering
    \includegraphics[width=0.95\linewidth]{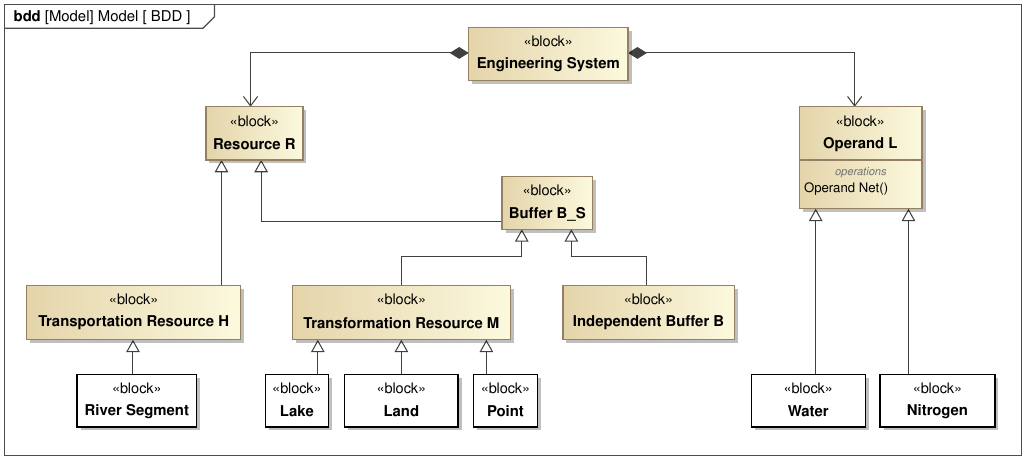}
    \caption{Block Definition Diagram that defines the lake, land, and river system form}
    \label{fig:hydro-bdd}
\end{figure}

\begin{figure}[htbp]
    \centering
    \includegraphics[width=0.95\linewidth]{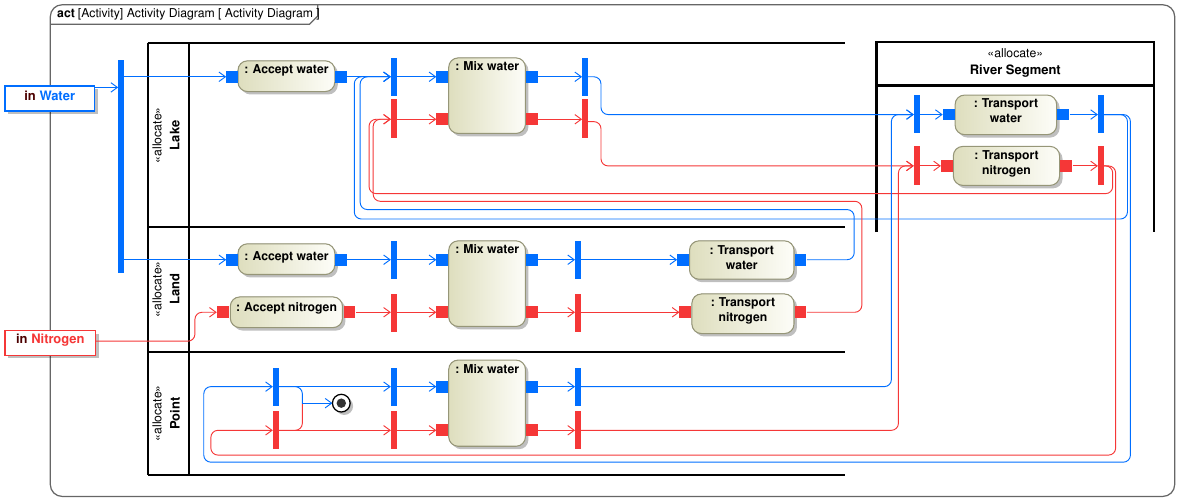}
    \caption{Activity Diagram that defines the lake, land, and river system function}
    \label{fig:hydro-act}
\end{figure}

The relationship between these three architectural levels is illustrated in Figures \ref{fig:LFESMetaArchitecturebdd}--\ref{fig:hydro-act}. Figures \ref{fig:LFESMetaArchitecturebdd} and \ref{fig:LFESMetaArchitectureact} present the HFGT meta-architecture, which defines the domain-independent ontology used throughout the methodology. Figures \ref{fig:hydro-bdd} and \ref{fig:hydro-act} specialize this ontology to hydrological systems, creating a hydrological reference architecture. In Section \ref{sec:instantiation}, this reference architecture is instantiated to represent the hydrological systems of the three illustrative examples.

Before introducing the individual modeling concepts, it is useful to define what HFGT considers an \emph{engineering system}. Throughout this paper, the term is used broadly to encompass natural, engineered, and socio-technical systems whose components interact to accomplish some function. The pedagogical hydrological system is an engineering system because its physical components interact to store, transport, and mix water and nitrogen.

\vspace{0.5em} \begin{defn}[Engineering System~\cite{De-Weck:2011:00}]\label{Defn:EngineeringSystem} 1) A class of systems characterized by a high degree of technical complexity, social intricacy, and elaborate processes aimed at fulfilling important functions in society, and 2) The term engineering systems is also used to refer to the engineering discipline that designs, analyzes, verifies, and validates \emph{engineering systems}.
\end{defn} \vspace{0.5em}

\subsubsection{Core Elements of HFGT}
The HFGT meta-architecture represents every engineering system using five fundamental concepts: operands, resources, processes, buffers, and capabilities. Each concept is introduced below using the pedagogical hydrological system.

\paragraph{System Operands}
Every engineering system acts upon something. In manufacturing, it may be parts. In electrical systems, it may be electricity. In transportation systems, it may be passengers or freight. HFGT refers to these entities as \emph{operands} $l_i \in L$.

\vspace{0.5em} \begin{defn}[System Operand~\cite{Walden:2015:00}]\label{def:operand}
An asset or object $l_i \in L$ that is operated on or consumed during the execution of a process.
\end{defn} \vspace{0.5em}

In the pedagogical hydrological system, the operands are water and nitrogen, $L=\{H_2O,N\}$. These materials move throughout the watershed, are stored within lakes and land segments, and are transported by runoff and river flow. These operands are defined within the HFGT Meta-Architecture in the Block Definition Diagram (BDD) in Fig. \ref{fig:hydro-bdd} and represented as operand flows in the Activity Diagram (ACT) of Fig. \ref{fig:hydro-act}.

\paragraph{System Processes}
Operands are transformed and transported throughout the system in what HFGT refers as \emph{processes} $p_w \in P$. 

\vspace{0.5em} \begin{defn}[System Process~\cite{Hoyle:1998:00,Walden:2015:00}]\label{def:process}
An activity $p \in P$ that transforms or transports a predefined set of input operands into a predefined set of outputs. 
\end{defn} \vspace{0.5em}

Examples include water as rainfall entering a lake, water as runoff leaving a land segment, water mixing within a lake, and water transport through a river. In manufacturing, assembly is a process; in transportation, movement of passengers is a process.

Processes are categorized into transformation processes which involve changes to operands within the system, and refined transportation processes, which include simultaneous movement and holding of operands. A transformation process includes the movement of an operand into or out of the system, such as the application of nitrogen from outside the system boundary or evaporation of water to outside the system boundary. Water and nitrogen undergo a transportation process when they are moved along the river network.

The illustrative hydrological system represented here defines the set of processes $P$ to include the transformation processes 
$p_1$) accept water, $p_2$) accept nitrogen, and $p_3$) mix water, and the transportation processes $p_4$) transport water and $p_5$) transport nitrogen.
These processes are represented in the Activity Diagram (ACT) of Fig. \ref{fig:hydro-act}.

\paragraph{System Resources} \label{subsubsec:resources}
The processes that transform and transport operands are facilitated by physical components of the system, which HFGT represents as \emph{resources}. Resources are the physical or logical entities that perform processes within the system. 

\vspace{0.5em} \begin{defn}[System Resource~\cite{Walden:2015:00}]\label{def:resource}
An asset or object $r_v \in R$ that facilitates the execution of a process.  
\end{defn} \vspace{0.5em}

In the hydrological example, the resources are the land segments, lakes, rivers, and river points. In an electrical grid, generators, transmission lines, and substations would all be resources.

Resources are divided into transformation resources, independent buffers, and transportation resources. \emph{Buffers} are resources capable of storing or transforming operands at specific locations in space~\cite{Schoonenberg:2019:00, Farid:2022:ISC-J49}. 

\vspace{0.5em} \begin{defn}[Buffer~\cite{Schoonenberg:2019:00,Farid:2022:ISC-J49}]\label{def:buffer}
A resource $r_v \in R$ is a buffer $b_s \in B_S$ iff it is capable of storing or transforming one or more operands at a unique location in space.
\end{defn} \vspace{0.5em}

Lakes, land segments, and river points all can store water and nitrogen and are therefore buffers. River segments transport water but are not modeled as storage locations in this example. In manufacturing, inventory acts as a buffer; in energy systems, batteries are buffers.

While independent buffers are only capable of storing operands, transformation resources are capable of storing, transforming, and transporting operands. For example, land segments can accept nitrogen as fertilizer, store it, and transport it to lakes via runoff. 

For this hydrological example, resources $r_v \in R$ are further specified. Transformation resources $m \in M \subset B_S \subset R$ include lakes ($m_1$, $b_{s1}$, $r_1$), land segments ($m_2$, $b_{s2}$, $r_2$), and river points ($m_3$, $b_{s3}$, $r_3$). Transportation resources $h \in H \subset R$ include river segments ($h_1$, $r_4$).

HFGT also introduces operand--buffer pairs to distinguish the storage of different operands within the same buffer. An operand--buffer pair is a unique combination of an operand and the buffer that stores it, such that each buffer contains one storage place for each operand. These pairs provide the place representation when the system is formulated using two-dimensional incidence matrices rather than three-dimensional tensors. For the hydrological example, the set of operand-buffer places, $[l_i-b_s]_B \in [L-B_S]$, consists of the following sets of buffer-operand places:  
$[l_1-b_{s1}]_1$ H$_2$O--Lake, $[l_1-b_{s2}]_2$) H$_2$O--Land, $[l_1-b_{s3}]_3$) H$_2$O--Point, $[l_2-b_{s1}]_4$) N--Lake, $[l_2-b_{s2}]_5$) N--Land, and $[l_2-b_{s3}]_6$) N--Point.

These resources, including buffers, are defined within the HFGT Meta-Architecture in the Block Definition Diagram (BDD) in Fig. \ref{fig:hydro-bdd} and the Activity Diagram (ACT) of Fig. \ref{fig:hydro-act}.

\paragraph{Capabilities} \label{subsubsec:capabilities}
Different resources can perform different sets of processes. HFGT represents these allowable actions as capabilities.

\vspace{0.5em} \begin{defn}[Capability~\cite{Schoonenberg:2019:00,Farid:2022:ISC-J49,Farid:2016:ISC-BC06}]\label{def:capability}
An action $e_{wv} \equiv e_{\psi} \in {\mathcal E}_S$ (in the SysML sense) defined by a system process $p_w \in P$ being executed by a resource $r_v \in R$.  
\end{defn} \vspace{0.5em}

A lake can receive rainfall, store water, mix constituents, and discharge flow. A river transports water but does not receive fertilizer directly. Consequently, these resources possess different capabilities. 

Capabilities distinguish what a resource \textbf{can} do from what it is doing at any specific time. For example, land segments as modeled here \textbf{can} receive fertilizer as inputs but a given instantiated land segment might not (e.g., if the land segment is not used for agricultural or other nitrogen-producing purposes). 

The capabilities of the hydrological system, $\epsilon_\psi \in {\mathcal E}_S$, concatenate the following sets of capability types: $\epsilon_1$) lake accepts H$_2$O, $\epsilon_2$) land accepts H$_2$O, $\epsilon_3$) lake accepts N, $\epsilon_4$) land accept N, $\epsilon_5$) lake mixes H$_2$O, $\epsilon_6$) land mixes H$_2$O, $\epsilon_7$) point mixes H$_2$O, $\epsilon_8$) land transports H$_2$O, $\epsilon_9$) land transports N, $\epsilon_{10}$) river transports H$_2$O, and $\epsilon_{11}$) river transports N. 

These capabilities are represented in the Activity Diagram (ACT) of Fig. \ref{fig:hydro-act} where a capability is the process allocated to a given resource.

\subsection{Mathematical Representation of HFGT} \label{sec:hfgtmath}
The preceding section established the conceptual ontology of engineering systems through resources, processes, operands, and capabilities. This section translates that ontology into a mathematical representation suitable for simulation and optimization. HFGT accomplishes this transformation through hetero-functional incidence tensors, Engineering System Nets, and the Hetero-functional Network Minimum Cost Flow formulation. 

\subsubsection{Hetero-functional Incidence Tensors}
The HFGT ontology defines the possible relationships among system resources, processes, and operands. To enable computational analysis, these relationships must be encoded mathematically. The \textit{hetero-functional incidence tensors (HFITs)} provide this representation by mapping operands, buffers, and capabilities into a unified mathematical structure. Unlike traditional graph incidence matrices, which represent connectivity between homogeneous nodes and edges, HFITs capture the heterogeneity of engineering systems by explicitly representing which operands are acted upon by which capabilities at which buffer locations.

\vspace{0.5em} \begin{defn}[The Negative 3$^{rd}$ Order Hetero-functional Incidence Tensor (HFIT) $\widetilde{\mathcal M}_\rho^-$~\cite{Farid:2022:ISC-J49}]\label{Defn:D6}
The negative hetero-functional incidence tensor $\widetilde{\mathcal M_\rho}^- \in \{0,1\}^{|L|\times |B_S| \times |{\mathcal E}_S|}$  is a third-order tensor whose element $\widetilde{\mathcal M}_\rho^{-}(i,y,\psi)=1$ when the system capability ${\epsilon}_\psi \in {\mathcal E}_S$ pulls operand $l_i \in L$ from buffer $b_{s_y} \in B_S$.
\end{defn} \vspace{0.5em} 

\vspace{0.5em} \begin{defn}[The Positive  3$^{rd}$ Order Hetero-functional Incidence Tensor (HFIT)$\widetilde{\mathcal M}_\rho^+$~\cite{Farid:2022:ISC-J49}]
The positive hetero-functional incidence tensor $\widetilde{\mathcal M}_\rho^+ \in \{0,1\}^{|L|\times |B_S| \times |{\mathcal E}_S|}$  is a third-order tensor whose element $\widetilde{\mathcal M}_\rho^{+}(i,y,\psi)=1$ when the system capability ${\epsilon}_\psi \in {\mathcal E}_S$ injects operand $l_i \in L$ into buffer $b_{s_y} \in B_S$.
\end{defn} \vspace{0.5em}

The third-order HFITs are matricized to form the positive and negative hetero-functional incidence matrices:

\begin{equation}
M^{+},M^{-}\in\{0,1\}^{|L||B_S|\times |{\mathcal E}_S|}
\end{equation}

where each row corresponds to a unique operand-buffer pair (size given as $|L||B_S|$, the total number of operands times the total number of buffers) and each column corresponds to a system capability (size given as ${|\mathcal E}_S|$). The net incidence matrix is then defined as:

\begin{equation}
M=M^{+}-M^{-}.
\end{equation}

This transformation provides the mathematical representation of operand flows through the engineering system. Each column of $M$ describes how a capability changes the state of the system by removing operands from some buffers and adding them to others.

For our simple hydrological system, the operand-buffer pairs defined in Subsection \ref{subsubsec:resources} represent the row indices of the net incidence matrix. Likewise, the  capabilities defined in Subsection \ref{subsubsec:capabilities} represent the column indices. 


Therefore, the net incidence matrix, $M$, for the hydrological system takes a block matrix form:
\setlength{\arraycolsep}{0pt}
\begin{align}
M = \begin{bmatrix}
M^+_{1,1} & 0         & 0         & M_{1,4} & 0       & 0       & M^+_{1,7} & 0         & M_{1,9} & 0 \\
0         & M^+_{2,2} & 0         & 0       & M_{2,5} & 0       & M^-_{2,7} & 0         & 0        & 0 \\
0         & 0         & 0         & 0       & 0       & M_{3,6} & 0         & 0         & M_{3,9} & 0 \\
0         & 0         & 0         & M_{4,4} & 0       & 0       & 0         & M^+_{4,8} & 0        & M_{4,10} \\
0         & 0         & M^+_{5,3} & 0       & M_{5,5} & 0       & 0         & M^-_{5,8} & 0        & 0 \\
0         & 0         & 0         & 0       & 0       & M_{6,6} & 0         & 0         & 0        & M_{6,10} 
\end{bmatrix} \label{blockmatrixform}
\end{align}
\setlength{\arraycolsep}{5pt} 
The convention $M_{B,\psi}$ is adopted to reflect an incidence matrix block corresponding to the operand-buffer pair row index $B$ and the $\psi$ capability index. 






\subsubsection{Engineering System Dynamics}
The hetero-functional incidence matrices define the structural relationships among operands, buffers, and capabilities. To represent the temporal evolution of an engineering system, HFGT formulates these relationships as an Engineering System Net (ESN), a Petri-net representation that tracks operand storage and capability execution through discrete time.

\vspace{0.5em} 
\begin{defn}[Engineering System Net~\cite{Schoonenberg:2022:00}]\label{Defn:ESN}
An elementary Petri net ${\mathcal N} = \{S, {\mathcal E}_S, \textit{M}, W, Q\}$, where
\begin{itemize}[noitemsep]
    \item $S$ is the set of places with size: $|L||B_S|$,
    \item ${\mathcal E}_S$ is the set of transitions with size: $|{\mathcal E}|$,
    \item $\textit{M}$ is the set of arcs, with the associated incidence matrices: $M = M^+ - M^-$,
    \item $W$ is the set of weights on the arcs, as captured in the incidence matrices,
    \item $Q=[Q_B; Q_E]$ is the marking vector for both the set of places and the set of transitions. 
\end{itemize}
\end{defn} 

The \textit{Engineering System Net State Transition Function }is equivalent to a generalized conservation equation. Specifically, the state of both the set of places and the set of transitions evolves according to the difference between operand additions and removals caused by executing system capabilities. 

\vspace{0.5em} \begin{defn}[Engineering System Net State Transition Function~\cite{Schoonenberg:2022:00}]\label{Defn:ESN-STF}
The  state transition function of the engineering system net $\Phi()$ is:
\begin{equation}\label{CH6:eq:PhiCPN}
Q[k+1]=\Phi(Q[k],U^-[k], U^+[k]) \quad \forall k \in \{1, \dots, K\}
\end{equation}
where $k$ is the discrete time index, $K$ is the simulation time horizon, $Q=[Q_{B}; Q_{\mathcal E}]$, $Q_B$ has size $|L||B_S| \times 1$, $Q_{\mathcal E}$ has size $|{\mathcal E}_S|\times 1$, the input firing vector $U^-[k]$ has size $|{\mathcal E}_S|\times 1$, and the output firing vector $U^+[k]$ has size $|{\mathcal E}_S|\times 1$.  
\begin{align}\label{CH6:CH6:eq:Q_B:HFNMCFprogram}
Q_{B}[k+1]&=Q_{B}[k]+{M}^+U^+[k]\Delta T-{M}^-U^-[k]\Delta T \\ \label{CH6:CH6:eq:Q_E:HFNMCFprogram}
Q_{\mathcal E}[k+1]&=Q_{\mathcal E}[k]-U^+[k]\Delta T +U^-[k]\Delta T
\end{align}
where $\Delta T$ is the duration of the simulation time step.  
\end{defn} 
\vspace{1em}

For hydrological systems, Equation \ref{CH6:CH6:eq:Q_B:HFNMCFprogram} corresponds directly to a mass balance, where buffer storage values in $Q_B$ represent water volumes or constituent nitrogen masses, and capability firings in $U$ represent flows of water or nitrogen between storage locations. Equation \ref{CH6:CH6:eq:Q_E:HFNMCFprogram} would represent storage within capabilities themselves $Q_E$. This storage is determined by a delay occurring between the time a process begins and ends, e.g., between start of the transportation of water from a lake to when it arrives at a river point. For our simple hydrological system, this delay is treated as instantaneous such that the input and output firing vectors $U^+$ and $U^-$ are equivalent.

Based on the defined set of operand-buffer places in Subsection \ref{subsubsec:resources}, the storage of water and nitrogen at operand buffer locations are represented by the following vertically concatenated vectors: 
\begin{align}
Q_B= 
\begin{array}{c}
[Q_{H_2O-Lake}\:\:;\:\: Q_{H_2O-Land}\:\:;\:\: Q_{H_2O-Point}\:\:;\:\: \\ Q_{N-Lake}\:\:;\:\: Q_{N-Land}\:\:;\:\: Q_{N-Point}]
\end{array}
\end{align}
The notation $Q_{H_2O}$ is introduced to reflect the first three elements of $Q_B$ associated with the operand water while $Q_N$ is introduced to reflect the last three elements of $Q_B$ associated with the operand nitrogen.  $Q_{H_2O}$ has volumetric units while $Q_N$ has mass units.  Next, the transitions of the engineering system net $U$ must reflect the different types of capabilities defined.  Consequently, $U$ has a structure composed of ten vertically concatenated vectors. 
\begin{align}
U=\begin{array}{c}
[U_{AcceptH_2O-Lake}\:\:;\:\: U_{AcceptH_2O-Land}\:\:;\:\: U_{AcceptN-Land} \:\:;\:\: \\
U_{MixH_2O-Lake} \:\:;\:\: U_{MixH_2O-Land} \:\:;\:\: U_{MixH_2O-Point} \:\:;\:\: \\
U_{TranspH_2O-Land} \:\:;\:\: U_{TranspN-Land} \:\:;\:\: \\ U_{TranspH_2O-River} \:\:;\:\: U_{TranspN-River}]
\end{array}
\end{align}


In some application domains, most notably production systems~\cite{Schoonenberg:2017:01,Farid:2008:IEM-J04,Farid:2008:IEM-J05} and healthcare systems~\cite{Khayal:2021:00,Khayal:2018:00,Khayal:2015:ISC-J20} respectively have products and patients as operands with often very complex operand state evolution. HFGT represents these dynamics using Operand Nets. An Operand Net describes the evolution of an individual operand through its own state variables and transitions.

\vspace{0.5em} \begin{defn}[Operand Net~\cite{Farid:2008:IEM-J04,Schoonenberg:2019:00,Khayal:2018:00,Schoonenberg:2017:01}]\label{Defn:OperandNet} Given operand $l_i$, an elementary Petri net ${\mathcal N}_{l_i}= \{S_{l_i}, {\mathcal E}_{l_i}, \textit{M}_{l_i}, W_{l_i}, Q_{l_i}\}$ where 
\begin{itemize}[noitemsep]
\item $S_{l_i}$ is the set of places describing the operand's state.  
\item ${\mathcal E}_{l_i}$ is the set of transitions describing the evolution of the operand's state.
\item $\textit{M}_{l_i} \subseteq (S_{l_i} \times {\mathcal E}_{l_i}) \cup ({\mathcal E}_{l_i} \times S_{l_i})$ is the set of arcs, with the associated incidence matrices: $M_{l_i} = M^+_{l_i} - M^-_{l_i} \quad \forall l_i \in L$.  
\item $W_{l_i} : \textit{M}_{l_i}$ is the set of weights on the arcs, as captured in the incidence matrices $M^+_{l_i},M^-_{l_i} \quad \forall l_i \in L$.  
\item $Q_{l_i}= [Q_{Sl_i}; Q_{{\mathcal E}l_i}]$ is the marking vector for both the set of places and the set of transitions. 
\end{itemize}
\end{defn} 

\begin{defn}[Operand Net State Transition Function~\cite{Farid:2008:IEM-J04,Schoonenberg:2019:00,Khayal:2018:00,Schoonenberg:2017:01}]\label{Defn:OperandNet-STF}
The  state transition function of each operand net $\Phi_{l_i}()$ is:
\begin{align} \label{CH6:eq:PhiSPN}
Q_{l_i}[k+1]=\Phi_{l_i}(Q_{l_i}[k],U_{l_i}^-[k], U_{l_i}^+[k]) \quad \forall \begin{array}{c} k \in \{1, \dots, K\} \\ i \in \{1, \dots, L\}
\end{array} 
\end{align}
where $Q_{l_i}=[Q_{Sl_i}; Q_{{\mathcal E} l_i}]$, $Q_{Sl_i}$ has size $|S_{l_i}| \times 1$, $Q_{{\mathcal E} l_i}$ has size $|{\mathcal E}_{l_i}| \times 1$, the input firing vector $U_{l_i}^-[k]$ has size $|{\mathcal E}_{l_i}|\times 1$, and the output firing vector $U^+[k]$ has size $|{\mathcal E}_{l_i}|\times 1$.  

\begin{align}\label{X}
Q_{Sl_i}[k+1]&=Q_{Sl_i}[k]+{M_{l_i}}^+U_{l_i}^+[k]\Delta T - {M_{l_i}}^-U_{l_i}^-[k]\Delta T \\ \label{CH6:CH eq:Q_E:HFNMCFprogram}
Q_{{\mathcal E} l_i}[k+1]&=Q_{{\mathcal E} l_i}[k]-U_{l_i}^+[k]\Delta T +U_{l_i}^-[k]\Delta T
\end{align}
\end{defn} \vspace{0.5em}

Such operand state behavior is predicated on a Lagrangian view rather than Eulerian view of the system~\cite{White:1994:00}.  Therefore, water system models -- which adopt an Eulerian view in which water and nutrient states are associated with spatial control volumes rather than individual operand trajectories -- do not make use of operand nets and their state transition functions. Consequently, the hydrological application considered here only utilizes the Engineering System Net.

\subsubsection{Hetero-functional Network Minimum Cost Flow Formulation}

The Engineering System Net provides a dynamic representation of system evolution through operand storage and capability execution. The Hetero-functional Network Minimum Cost Flow (HFNMCF) formulation extends this representation into a constrained optimization framework that determines feasible system behavior while satisfying conservation laws, capability relationships, boundary conditions, and physical device models. This section defines the formulation of the HFNMCF problem while deriving the domain-specific formulation for the described hydrological systems.  

The general HFNMCF formulation is expressed as:
\begin{subequations}\label{eq:simplehydro}
\begin{align}\label{Eq:ObjFunc}
Z &= \sum_{k=1}^{K-1} x^T[k] F_{QP} x[k] + f_{QP}^T x[k]
\end{align}
\begin{align} \label{Eq:ESN-STF1}
 -Q_{B}[k+1]+Q_{B}[k]+{M}^+U^+[k]\Delta T - {M}^-U^-[k]\Delta T=&0 \\  \label{Eq:ESN-STF2}
-Q_{\mathcal E}[k+1]+Q_{\mathcal E}[k]-U^+[k]\Delta T + U^-[k]\Delta T=&0 \\
\label{Eq:DurationConstraint}
- U_\psi^+[k + k_{d\psi}] + U_\psi^-[k] = 
& 0  \\ \label{Eq:OperandNet-STF1}
-Q_{Sl_i}[k+1]+Q_{Sl_i}[k]+{M}_{l_i}^+U_{l_i}^+[k]\Delta T - {M}_{l_i}^-U_{l_i}^-[k]\Delta T=&0  \\ 
\label{Eq:OperandNet-STF2}
-Q_{{\mathcal E}l_i}[k+1]+Q_{{\mathcal E}l_i}[k]-U_{l_i}^+[k]\Delta T + U_{l_i}^-[k]\Delta T=&0 \\ 
\label{Eq:OperandNetDurationConstraint}
- U_{xl_i}^+[k+k_{dxl_i}]+ U_{xl_i}^-[k] = &0 \\  
\label{Eq:SyncPlus}
U^+_L[k] - \widehat{\Lambda}^+ U^+[k] =&0 \\ \label{Eq:SyncMinus}
U^-_L[k] - \widehat{\Lambda}^- U^-[k] =&0 \\ \label{CH6:eq:HFGTprog:comp:Bound}
\begin{bmatrix}
D_{Up} & \mathbf{0} \\ \mathbf{0} & D_{Un}
\end{bmatrix} \begin{bmatrix}
U^+ \\ U^-
\end{bmatrix}[k] = \begin{bmatrix}
C_{Up} \\ C_{Un}
\end{bmatrix}[k] \\ \label{Eq:OperandRequirements}
\begin{bmatrix}
E_{Lp} & \mathbf{0} \\ \mathbf{0} & E_{Ln}
\end{bmatrix} \begin{bmatrix}
U^+_{l_i} \\ U^-_{l_i}
\end{bmatrix}[k] = \begin{bmatrix}
F_{Lpi} \\ F_{Lni}
\end{bmatrix}[k] \! \\  
\label{CH6:eq:HFGTprog:comp:Init} 
\begin{bmatrix} Q_B ; Q_{\mathcal E} ; Q_{SL} \end{bmatrix}[1] =\begin{bmatrix} C_{B1} ; C_{{\mathcal E}1} ; C_{{SL}1} \end{bmatrix} \\ \label{CH6:eq:HFGTprog:comp:Fini}
\begin{bmatrix} Q_B ; Q_{\mathcal E} ; Q_{SL} ; U^- ; U_L^- \end{bmatrix}[K+1] =   \begin{bmatrix} C_{BK} ; C_{{\mathcal E}K} ; C_{{SL}K} ; \mathbf{0} ; \mathbf{0} \end{bmatrix}\\ \label{ch6:eq:QPcanonicalform:3}
\underline{E}_{CP} \leq D(X) \leq \overline{E}_{CP} \\ \label{Eq:DeviceModels}
g(X,Y) =&0  \\ \label{Eq:DevicModels2}
h(Y) \leq&0 
\\ \notag  \begin{array}{c}
  \forall k \in \{1, \dots, K\}, \quad \psi \in \{1,\dots,\mathcal{E}_S\}, \quad i \in \{1, \dots, |L|\}, \\ x\in \{1, \dots, |{\mathcal E}_{l_i}\}, \quad l_i \in \{1, \dots, |L|\}
\end{array} 
\end{align}
\end{subequations}
where $X=\left[x[1]; \ldots; x[K]\right]$  is the vector of primary decision variables and $Y=\left[y[1]; \ldots; y[K]\right]$  is the vector of auxiliary decision variables at time $k$. 

\vspace{1em}
For the illustrative hydrological system, the general form of the HFNMCF optimization problem in \ref{Eq:ObjFunc}-\ref{Eq:DevicModels2} simplifies to:
\begin{subequations}
\begin{alignat}{3}\label{eq:simplified-objective}
\text{minimize }Z &= 0  \\ 
\label{Eq:simplified-STF1}
-Q_{B}[k+1]+Q_{B}[k]+{M}U[k]\Delta T &= 0 \\ 
\label{Eq:simplified-exogenous}
D_{U}U[k] &= C_{U}[k]  \\
\label{Eq:simplified-initCond} 
Q_B[1] &= C_{B1} \\
\label{Eq:simplified-resistance-land}
U_{TranspH_2O-Land}[k] - R_{TranspH_2O-Land}^{-1}(\rho g)&\times \notag \\ (-M_{TranspH_2O-Land}^{T})\left(A_{H_2O}^{-1}\left(Q_{H_2O}-Q_{H_2O,min}\right)+z_{B}\right) &=0  \\
\label{Eq:simplified-resistance-river}
 U_{TranspH_2O-River}[k] - R_{TranspH_2O-River}^{-1}(\rho g)& \times \notag \\ (-M_{TranspH_2O-River}^{T})\left(A_{H_2O}^{-1}\left(Q_{H_2O}-Q_{H_2O,min}\right)+z_{B}\right) &=0 \\
Q_{N-Land}[k] \cdot M^-_{2,7} \times U_{TranspH_2O-Land}[k]\Delta T  -  & \notag\\  Q_{H_2O-Land}[k] \cdot M^-_{5,8} \times U_{TranspN-Land}[k]\Delta T &= 0 \label{Eq:simplified-mixing-land} \\
Q_{N-Lake}[k] \cdot M^-_{1,9} \times U_{TranspH_2O-River}[k]\Delta T -  & \notag\\  Q_{H_2O-Lake}[k] \cdot M^-_{4,10} \times U_{TranspN-River}[k]\Delta T &= 0 \label{Eq:simplified-mixing-lake}\\
Q_{N-Point}[k] \cdot M^-_{3,9} \times U_{TranspH_2O-River}[k]\Delta T -  & \notag\\ Q_{H_2O-Point}[k] \cdot M^-_{6,10} \times U_{TranspN-River}[k]\Delta T &= 0  \label{Eq:simplified-mixing-point}\\
\begin{array}{c}
     \forall k \in \{1, \dots, K\} 
\end{array} \nonumber
\end{alignat}
\end{subequations}

\paragraph{Objective Function}
In Eq.  \ref{Eq:ObjFunc}, $Z$ is a convex objective function separable in discrete time steps $k$.  Matrix $F_{QP}$ and vector $f_{QP}$ allow quadratic and linear costs to be incurred from the place and transition markings in both the engineering system net and operand nets. 
\begin{itemize}[noitemsep]
\item $F_{QP}$ is a positive semi-definite, diagonal, quadratic coefficient matrix.
\item $f_{QP}$ is a linear coefficient matrix.
\end{itemize}

For the hydrological example, the objective function in Equation \ref{Eq:ObjFunc} simplifies to Equation \ref{eq:simplified-objective} because the simulation of a watershed system is an initial value problem rather than an optimization problem. That is to say, no operand flows, storage, or proxy variables are optimized in this example.

\paragraph{Equality Constraints}
The equality constraints of the general HFNMCF formulation (Eq. \ref{Eq:ESN-STF1}--\ref{CH6:eq:HFGTprog:comp:Fini}) and their simplification for the pedagogical hydrological system (Eq. \ref{Eq:simplified-STF1}--\ref{Eq:simplified-initCond}) are summarized as follows:
\begin{itemize}[noitemsep]
    \item Equations \ref{Eq:ESN-STF1} and \ref{Eq:ESN-STF2} describe the \textbf{Engineering System Net State Transition Function} (Defn \ref{Defn:ESN} \& \ref{Defn:ESN-STF}), and Equation \ref{Eq:DurationConstraint} defines \textbf{Engineering System Net State Transition Function} duration constraint where the end of the $\psi^{th}$ transition occurs $k_{d\psi}$ time steps after its beginning. For the hydrological system, Equations \ref{Eq:ESN-STF1}--\ref{Eq:DurationConstraint} simplify to Equation \ref{Eq:simplified-STF1} given that: 
    \begin{itemize}
        \item $\Delta T$ is assumed to be sufficiently short so that the discretization of time does not affect the continuous-time dynamics of the watershed system.  
        \item Therefore, each transition is assumed to occur instantaneously.  $k_{d\psi} = 0$.  As a result, $U^+[k] = U^-[k] = U[k]\:\forall k$.
        \item Additionally, Equation \ref{Eq:ESN-STF2} collapses to triviality.  
        \item Moreover, the positive and negative matrices of the hetero-functional incidence matrix can be combined: $M = M^+ - M^-$.  
        \item Finally, $Q_B[k]$ tracks the storage of water volume and nitrogen mass at each buffer location.
\end{itemize}
\item Equations \ref{Eq:OperandNet-STF1} and \ref{Eq:OperandNet-STF2} describe the \textbf{Operand Net State Transition Function} ${\mathcal N}_{l_i}$ (Defn. \ref{Defn:OperandNet} \& \ref{Defn:OperandNet-STF}) associated with each operand $l_i \in L$, and Equation \ref{Eq:OperandNetDurationConstraint} defines \textbf{Operand Net State Transition Function} duration constraint where the end of the $x^{th}$ transition occurs $k_{dx_{l_i}}$ time steps after its beginning. Correspondingly, Equations \ref{Eq:SyncPlus} and \ref{Eq:SyncMinus} are synchronization constraints that couple the input and output firing vectors of the \textbf{Engineering System Net} to the input and output firing vectors of the \textbf{Operand Net} respectively. $U_L^-$ and $U_L^+$ are the vertical concatenations of the input and output firing vectors $U_{l_i}^-$ and $U_{l_i}^+$ respectively.
\begin{align}
U_L^-[k]&=\left[U^-_{l_1}; \ldots; U^-_{l_{|L|}}\right][k] \\
U_L^+[k]&=\left[U^+_{l_1}; \ldots; U^+_{l_{|L|}}\right][k]
\end{align}
Equations \ref{Eq:OperandNet-STF1}--\ref{Eq:SyncMinus} are not needed for the hydrological system given that the two operands, water and nitrogen, do not undergo state changes in this system, and therefore do not require operand nets to track state changes. 
\item Equations \ref{CH6:eq:HFGTprog:comp:Bound} and \ref{Eq:OperandRequirements} are boundary conditions.  Eq. \ref{CH6:eq:HFGTprog:comp:Bound} is a boundary condition constraint that allows some of the \textbf{Engineering System Net} firing vectors decision variables to be set to an exogenous constant.  Eq. \ref{Eq:OperandRequirements} does the same for the \textbf{Operand Net} firing vectors. For the hydrological system, Equation \ref{CH6:eq:HFGTprog:comp:Bound} reduces to Equation \ref{Eq:simplified-exogenous}, capturing exogenous inputs such as precipitation or fertilizer. Equation \ref{Eq:OperandRequirements} also pertains to operand nets and is therefore not needed. 
\item Equations \ref{CH6:eq:HFGTprog:comp:Init} and \ref{CH6:eq:HFGTprog:comp:Fini} are the initial and final conditions of the \textbf{Engineering System Net} and the operand nets where $Q_{SL}$ is the vertical concatenation of the place marking vectors of the operand nets $Q_{Sl_i}$.
\begin{align}
Q_{SL}^-[k]&=\left[Q^-_{Sl_1}; \ldots; U^-_{Sl_{|L|}}\right][k] \\
U_{SL}^+[k]&=\left[U^+_{Sl_1}; \ldots; U^+_{Sl_{|L|}}\right][k]
\end{align}
For the hydrological system, the initial condition constraint simplifies to \ref{Eq:simplified-initCond}, and Equation \ref{CH6:eq:HFGTprog:comp:Fini} is not required because watershed system simulation is an initial value problem.  
\end{itemize}

\paragraph{Inequality Constraints}
 $D_{QP}()$ and vector $E_{QP}$ in Equation \ref{ch6:eq:QPcanonicalform:3} place capacity constraints at each time step on this vector of primary decision variables:
 \begin{align*}
     x[k] = \begin{bmatrix} Q_B ; Q_{\mathcal E} ; Q_{SL} ; Q_{{\mathcal E}L} ; U^- ; U^+ ; U^-_L ; U^+_L \end{bmatrix}[k] \quad \forall k \in \{1, \dots, K\}.
 \end{align*}

The hydrological system defined in this manuscript does not require the use of inequality constraints.

\paragraph{Device Model Constraints}
g(X,Y) (Eq. \ref{Eq:DeviceModels}) and h(Y) (Eq. \ref{Eq:DevicModels2}) are a set of device model functions whose presence and nature depend on the specific problem application.  They can not be further elaborated until the application domain and its associated capabilities are identified. In the case of this paper, these device models include laws of resistance-controlled flow and assumption of uniform mixing of nitrogen within bodies of water.

Equation \ref{Eq:DeviceModels} concerning device model equality constraints is expanded to Eq. \ref{Eq:simplified-resistance-land}-\ref{Eq:simplified-mixing-point}.  Each of these equations is elaborated below.

\begin{itemize}[noitemsep]
\item Equations \ref{Eq:simplified-resistance-land} and \ref{Eq:simplified-resistance-river} are a matrix restatement of the familiar formula $R^{-1}P=\dot{V}$ where $R$ is the fluidic resistance, $P$ is the relevant pressure, and $\dot{V}$ is the associated volumetric flow rate.   Here, in order to express the constitutive law for fluidic resistance of water transport across land and river: 
\begin{itemize}[noitemsep]
    \item The transportation of water via the land and river have associated fluidic resistance values in the  $R_{TransportH_2O-Land}$ and $R_{TransportH_2O-River}$ matrices. They inversely control the rate of flow from the change in pressure as $R^{-1}$.
    \item Flow is driven by pressure differences proportional to the hydraulic head at each buffer place, and the term $\rho g$ converts the differences in head into pressure, $P$, where $\rho$ is the density of water, and $g$ is the gravitational acceleration constant.  
    \item The elements of the negative transpose of the hetero-functional incidence tensor, $(-M)^{T}_{TransportH_2O-Land}$ and $(-M)^{T}_{TransportH_2O-River}$, map pressure differences at buffer locations to their effluent transport processes.
    \item The pressure head, $P$, includes two parts: (1) a volume-dependent pressure term, derived from the effective water depth in the buffer, $A_{H_2O}^{-1}(Q_{H_2O} - Q_{H_2O,\min})$, where $A_B$ is the surface area of each buffer and $Q_{H_2O,\min}$ is the threshold volume below which flow does not occur, and (2) the elevation of the buffer, $z_B$. These terms account for the flow resulting from pressure differences in both water depth and elevation.
    \item These equations ultimately calculate the volumetric flow rate of water, $\dot{V}$, as the firing vectors for the transportation of water via the land, $U_{TransportH_2O-Land}$, and river, $U_{TransportH_2O-River}$.
\end{itemize}
\item Equations \ref{Eq:simplified-mixing-land}–\ref{Eq:simplified-mixing-point} are matrix statements that the nitrogen concentration of an outlet flow is equal to the nitrogen concentration of a lake, land segment, or river point.   Each of these device models has the familiar formula $\frac{m}{V}=\frac{\dot{m}}{\dot{V}}$, where the left hand side is the concentration of a lake, land segment, or river point and the right hand side is the concentration of its outlet flow. For computational stability, this familiar formula was reformulated to take the form $m\dot{V} = V\dot{m}$, eliminating the need for division operations. Furthermore, the notation $\times$ is a matrix product while $\cdot$ is the element-wise (Hadamard) product.  
\begin{itemize}[noitemsep]
    \item The terms $Q_{N-Land}$, $Q_{N-Lake}$, and $Q_{N-Point}$ represent the nitrogen mass, $m$, stored in each buffer class, while $Q_{H_2O-Land}$, $Q_{H_2O-Lake}$, and $Q_{H_2O-Point}$ represent the corresponding water volumes, $V$.  
    \item The negative Hetero-functional Incidence Tensor $M^-$ identifies the outflow processes for nitrogen and water that correspond to each land segment, lake, or point.
    \item The variables $U_{TransportN-Land}$ and $U_{TransportN-River}$ represent the nitrogen outflow, $\dot{m}$, via runoff and river transport, respectively. Likewise, $U_{TransportH_2O-Land}$ and $U_{TransportH_2O-River}$ represent the corresponding water outflows, $\dot{V}$
\end{itemize}
\end{itemize}

\section{Results: Instantiating the Hydrological System Examples} \label{sec:instantiation}

The hydrological system formulation described in Sections~\ref{sec:hfgtterms} and~\ref{sec:hfgtmath} enables the simulation of water and nutrient dynamics across a watershed while preserving mass balance, process structure, and physical constraints. The following subsections instantiate the methodology for three systems of increasing complexity. The first example is presented in detail to clarify each instantiation step, while the later examples emphasize the insights each one adds.

\subsection{Parameterization of the Illustrative Examples} \label{sec:parameters}
All three examples share one set of synthetic parameters, scaled to published regional values so that volumes, flows, and concentrations are physically plausible. Net water input was derived from the 1991--2020 normal annual precipitation for Blacksburg, Virginia (1083~mm~yr$^{-1}$)~\cite{NOAA:2021:00}, of which approximately 40\% was taken as water yield, the remainder being returned to the atmosphere by evapotranspiration in the humid eastern United States~\cite{Sanford:2013:00}. This input was distributed over four seasonal blocks (January--March, April--June, July--September, and October--December) with multipliers of 1.6, 1.1, 0.4, and 0.9, representing wetter winter--spring and drier summer--fall conditions. Lakes 1, 2, and 3 have surface areas of 30, 40, and 50~ha and threshold depths of 3.0, 4.0, and 3.5~m, below which water does not leave the lake ($Q_{H_2O,min} = A_{lake} \times$ threshold depth). In Examples 1 and 2, each lake receives the net water input from a contributing area 15 to 25 times its surface area; in Example 3, this contributing area is represented explicitly by the land segments. Nitrogen inputs to the land segments in Example 3 follow the Chesapeake Bay Program Phase~6 edge-of-stream total-nitrogen loading rates for full-season soybeans (31.1~kg~ha$^{-1}$~yr$^{-1}$), other agronomic crops (19.7~kg~ha$^{-1}$~yr$^{-1}$), and pasture (13.2~kg~ha$^{-1}$~yr$^{-1}$)~\cite{CBP:2018:00}, applied predominantly in spring (15, 55, 15, and 15\% by season). Initial lake total nitrogen in Examples 1 and 2 (0.6--1.2~mg~L$^{-1}$) exceeds the recommended lake criterion for the region (0.36~mg~L$^{-1}$)~\cite{EPA:2000:lakesIX}, so that these examples describe the flushing of nutrient-enriched lakes by low-nitrogen inflow.

River points represent storage within a river reach (5~km long and 20~m wide), and the most downstream point represents the receiving water body below the watershed outlet. Fluidic resistances were chosen to give each lake a response time of 20 days (60 days for a lake together with its land segment in Example 3), each land segment a response time of 30 days, and each river reach a 0.3~m head drop at mean flow. Elevations were then set so that every storage element sits at its target depth under mean flow, so that outflow is governed by storage rather than by fixed elevation differences. Initial storages were obtained from a multi-year spin-up of each system, and all simulations use a time step of $\Delta T = 6$~h over 52 weeks.

\subsection{Illustrative Example 1: A System with One Lake}
\begin{figure}[htbp]
    \centering
    \includegraphics[width=0.6\linewidth]{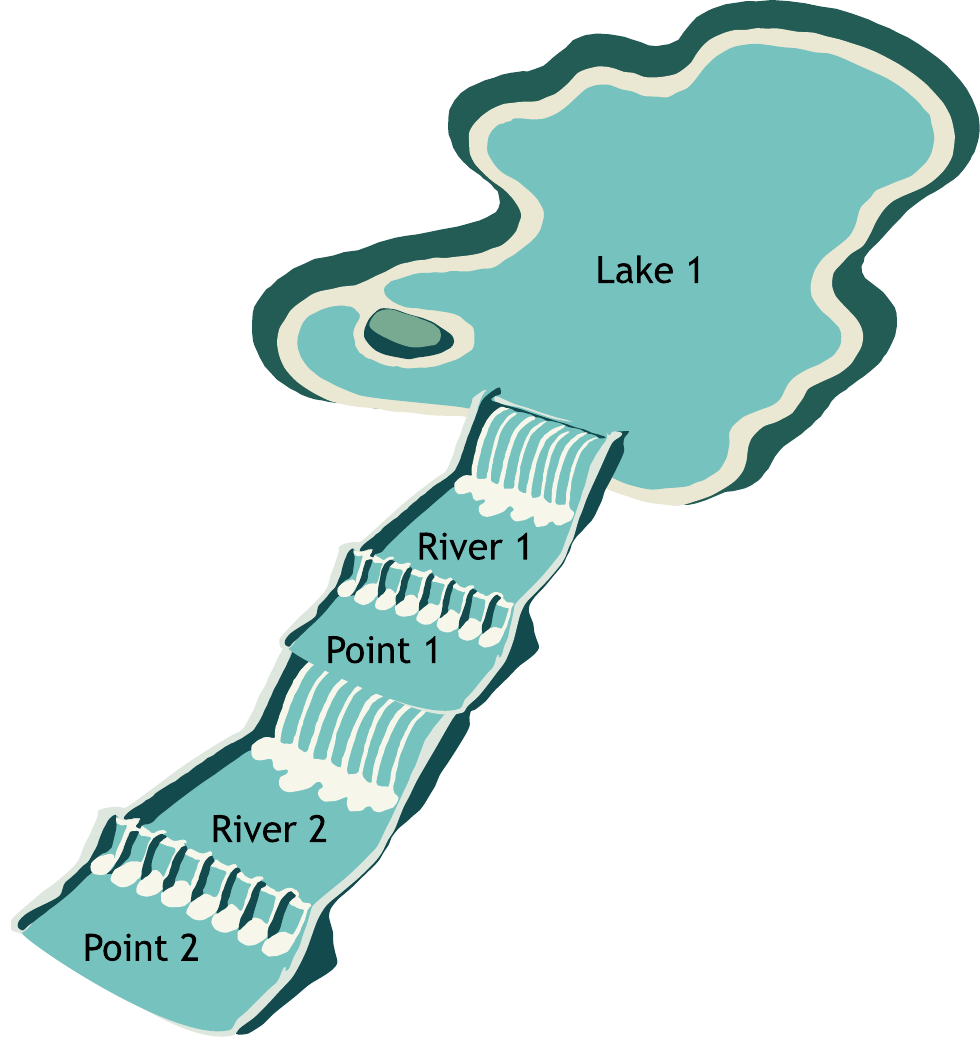}
    \caption{Diagram of the system with one lake: Lake 1 drains through River Segment 1 to Point 1, a river reach, and through River Segment 2 to Point 2, the receiving water below the watershed outlet.}
    \label{fig:diagramSingleLake}
\end{figure}

This illustrative system resembles a single CSTR configuration~\cite{kumar:2012:01}: one lake with an exogenous inflow and a resistance-controlled outflow. Based on the watershed reference architecture introduced in Section~\ref{sec:hfgtterms}, it is instantiated with one lake, two river points, and two river segments (Fig.~\ref{fig:diagramSingleLake}). River Segment 1 carries water and nitrogen from Lake 1 to Point 1, a river reach, and River Segment 2 carries them from Point 1 to Point 2, the receiving water below the outlet. Lake 1 and both points are transformation resources: each stores water and nitrogen and mixes them, and Lake 1 additionally accepts the exogenous water input.

\begin{figure}[htbp]
    \centering
    \includegraphics[width=0.9\linewidth]{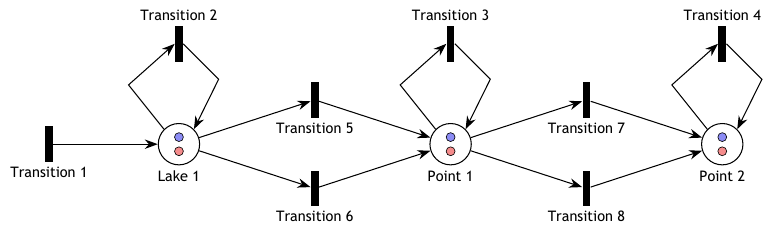}\\[2pt]
    \textbf{(A)}\\[10pt]
    \includegraphics[width=0.9\linewidth]{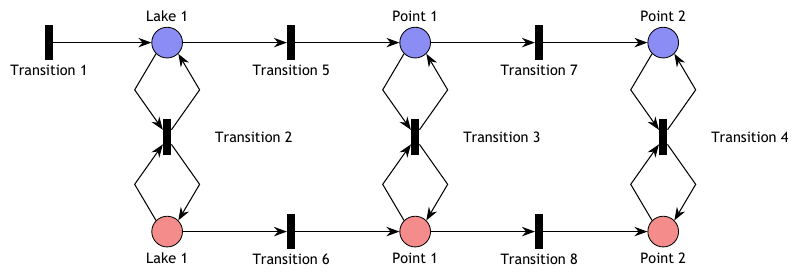}\\[2pt]
    \textbf{(B)}
    \caption{Petri net representations of the system with one lake: \textbf{(A)} colored Petri net and \textbf{(B)} elementary Petri net.}
    \label{fig:s1-both}
\end{figure}

The resulting architecture is first represented as a colored Petri net in Fig.~\ref{fig:s1-both}A, which captures the dynamics of water and nitrogen as they flow through the system. Each buffer (Lake 1, Point 1, and Point 2) is represented by a Petri net place, and water (blue) and nitrogen (red) are transformed within these places and transported between them by capabilities. The colored Petri net is then translated into the elementary Petri net shown in Fig.~\ref{fig:s1-both}B, which is necessary for modeling the system using matrix algebra. In the elementary Petri net, each buffer has a separate place for each operand, giving six buffer--operand places.

The eight capabilities in this system are:
\begin{enumerate}[noitemsep]
    \item Accept water at Lake 1 by Lake 1,
    \item Mix water and nitrogen at Lake 1 by Lake 1,
    \item Mix water and nitrogen at Point 1 by Point 1,
    \item Mix water and nitrogen at Point 2 by Point 2,
    \item Transport water from Lake 1 to Point 1 by River Segment 1,
    \item Transport nitrogen from Lake 1 to Point 1 by River Segment 1,
    \item Transport water from Point 1 to Point 2 by River Segment 2,
    \item Transport nitrogen from Point 1 to Point 2 by River Segment 2.
\end{enumerate}

This Petri net and its underlying architecture are encoded in an XML file and processed using the HFGT toolbox~\cite{Farid:2025:ISC-BKR01}, which generates the hetero-functional incidence tensors and associated metadata. These outputs are the inputs to a Julia-based simulation environment that advances the instantiated HFNMCF model in time.

The behavior of the system is governed by the general HFNMCF equations described in Section~\ref{sec:hfgtmath}, instantiated with matrices tailored to this configuration. The instantiated matrices corresponding to the mass balance (Eq.~\ref{Eq:simplified-STF1}), boundary conditions (Eq.~\ref{Eq:simplified-exogenous}), initial conditions (Eq.~\ref{Eq:simplified-initCond}), resistance-based flow (Eqs.~\ref{Eq:simplified-resistance-land}--\ref{Eq:simplified-resistance-river}), and complete mixing of nitrogen (Eqs.~\ref{Eq:simplified-mixing-land}--\ref{Eq:simplified-mixing-point}) are shown below.

\noindent The variables associated with buffer storage of operands, $Q_B$, are:
\begin{align*}
Q_B=& \begin{bNiceMatrix}
   Q_{H_2O} \\ Q_{N}
\end{bNiceMatrix} = \begin{bNiceMatrix}
    V_{lake1} \\ V_{point1} \\ V_{point2} \\ m_{lake1} \\ m_{point1} \\ m_{point2}
    \end{bNiceMatrix}, \quad C_{B1}= \begin{bNiceMatrix}
     V_{lake1,0} \\ V_{point1,0} \\ V_{point2,0} \\ m_{lake1,0} \\ m_{point1,0} \\ m_{point2,0}
    \end{bNiceMatrix}
\end{align*}
The variables associated with buffer storage of water, $Q_{H_2O}$, are:
\begin{align*}
    Q_{H_2O}=& \begin{bNiceMatrix}
    V_{lake1} \\ V_{point1} \\ V_{point2}
    \end{bNiceMatrix}, \:\: Q_{H_2O,min}=\begin{bNiceMatrix}
     V_{lake1,min} \\ V_{point1,min} \\ V_{point2,min}
    \end{bNiceMatrix}, \:\:  A_{H_2O}^{-1}= \begin{bNiceMatrix}
        \frac{1}{A_{lake1}} & 0 & 0 \\
        0 & \frac{1}{A_{point1}} & 0 \\
        0 & 0 & \frac{1}{A_{point2}}
    \end{bNiceMatrix}
\end{align*}
\begin{align*}
    Q_{H_2O-land}=& \begin{bNiceMatrix}
    \text{N/A}
    \end{bNiceMatrix}, \:\:
    Q_{H_2O-lake}= \begin{bNiceMatrix}
    V_{lake1}
    \end{bNiceMatrix}, \:\:
    Q_{H_2O-point}= \begin{bNiceMatrix}
    V_{point1} \\ V_{point2}
    \end{bNiceMatrix}
\end{align*}
The variables associated with buffer storage of nitrogen, $Q_{N}$, are:
\begin{align*}
Q_N =&  \begin{bNiceMatrix}
    m_{lake1} \\ m_{point1} \\ m_{point2}
    \end{bNiceMatrix}, \:\:
    Q_{N-land}= \begin{bNiceMatrix}
    \text{N/A}
    \end{bNiceMatrix}, \:\:
    Q_{N-lake}= \begin{bNiceMatrix}
    m_{lake1}
    \end{bNiceMatrix}, \:\:
    Q_{N-point}= \begin{bNiceMatrix}
    m_{point1} \\ m_{point2}
    \end{bNiceMatrix}
\end{align*}
The elevation attributes of the water places are:
\begin{align*}
z_{B} =\begin{bNiceMatrix}
     z_{lake1} \\ z_{point1} \\ z_{point2}
    \end{bNiceMatrix}
\end{align*}
The variables associated with process flow rates, $U$, are:
\begin{align*}
U=& \begin{bNiceMatrix}
        \dot{V}_{1} \\ \dot{T}_{2} \\ \dot{T}_{3} \\ \dot{T}_{4} \\ \dot{V}_{5} \\ \dot{m}_{6} \\ \dot{V}_{7} \\ \dot{m}_{8}
    \end{bNiceMatrix}, \quad C_U = \begin{bNiceMatrix}
        \dot{V}_{1}^\ddagger \\ 0 \\ 0 \\ 0 \\ 0 \\ 0 \\ 0 \\ 0
    \end{bNiceMatrix}, \quad
D_U = \begin{bNiceMatrix}
        1 & 0 & 0 & 0 & 0 & 0 & 0 & 0\\
        0 & 0 & 0 & 0 & 0 & 0 & 0 & 0\\
        0 & 0 & 0 & 0 & 0 & 0 & 0 & 0\\
        0 & 0 & 0 & 0 & 0 & 0 & 0 & 0\\
        0 & 0 & 0 & 0 & 0 & 0 & 0 & 0\\
        0 & 0 & 0 & 0 & 0 & 0 & 0 & 0\\
        0 & 0 & 0 & 0 & 0 & 0 & 0 & 0\\
        0 & 0 & 0 & 0 & 0 & 0 & 0 & 0
    \end{bNiceMatrix}
\end{align*}
where $\dot{V}_{1}^\ddagger$ is the exogenous net water input to Lake 1. The variables associated with the transport processes of water are:
\begin{align*}
U_{TranspH_2O-Land} &= \begin{bNiceMatrix}\text{N/A} \end{bNiceMatrix},
&&U_{TranspH_2O-River} = \begin{bNiceMatrix} \dot{V}_{5} \\ \dot{V}_{7} \end{bNiceMatrix}, \\
R_{TranspH_2O-Land}^{-1} &= \begin{bNiceMatrix}\text{N/A}\end{bNiceMatrix},
&&R_{TranspH_2O-River}^{-1} = \begin{bNiceMatrix} \frac{1}{R_5} & 0 \\ 0 & \frac{1}{R_7} \end{bNiceMatrix}, \\
-M^T_{TranspH_2O-Land} &= \begin{bNiceMatrix}\text{N/A}\end{bNiceMatrix},
&-&M^T_{TranspH_2O-River} = \begin{bNiceMatrix} 1 & -1 & 0 \\ 0 & 1 & -1 \end{bNiceMatrix}
\end{align*}
The variables associated with the transport processes of nitrogen are:
\begin{align*}
U_{TranspN-Land} = \begin{bNiceMatrix}\text{N/A}\end{bNiceMatrix}, \quad
    U_{TranspN-River} = \begin{bNiceMatrix}
        \dot{m}_{6} \\ \dot{m}_{8}
    \end{bNiceMatrix}
\end{align*}
The hetero-functional incidence matrices, with rows ordered as in $Q_B$ and columns as in $U$, are:
\begin{align*}
M^{+}= \begin{bNiceMatrix}
        1 & 1 & 0 & 0 & 0 & 0 & 0 & 0 \\
        0 & 0 & 1 & 0 & 1 & 0 & 0 & 0 \\
        0 & 0 & 0 & 1 & 0 & 0 & 1 & 0 \\
        0 & 1 & 0 & 0 & 0 & 0 & 0 & 0 \\
        0 & 0 & 1 & 0 & 0 & 1 & 0 & 0 \\
        0 & 0 & 0 & 1 & 0 & 0 & 0 & 1
    \end{bNiceMatrix}, \:\:
    M^{-}=& \begin{bNiceMatrix}
        0 & 1 & 0 & 0 & 1 & 0 & 0 & 0 \\
        0 & 0 & 1 & 0 & 0 & 0 & 1 & 0 \\
        0 & 0 & 0 & 1 & 0 & 0 & 0 & 0 \\
        0 & 1 & 0 & 0 & 0 & 1 & 0 & 0 \\
        0 & 0 & 1 & 0 & 0 & 0 & 0 & 1 \\
        0 & 0 & 0 & 1 & 0 & 0 & 0 & 0
    \end{bNiceMatrix} \\
    M^{+} - M^{-} =M =& \begin{bNiceMatrix}
        1 & 0 & 0 & 0 & -1 & 0 & 0 & 0 \\
        0 & 0 & 0 & 0 & 1 & 0 & -1 & 0 \\
        0 & 0 & 0 & 0 & 0 & 0 & 1 & 0 \\
        0 & 0 & 0 & 0 & 0 & -1 & 0 & 0 \\
        0 & 0 & 0 & 0 & 0 & 1 & 0 & -1 \\
        0 & 0 & 0 & 0 & 0 & 0 & 0 & 1
    \end{bNiceMatrix}
\end{align*}
Each transport column of $M$ removes an operand from its origin ($-1$) and adds it to its destination ($+1$), and the three mixing columns sum to zero, so the formulation conserves water and nitrogen by construction. The submatrices of the negative hetero-functional incidence matrix through which water and nitrogen leave buffers are:
\begin{align*}
\begin{array}{c}
    M^-_{2,7}=\begin{bNiceMatrix}\text{N/A} \end{bNiceMatrix}, \quad
    M^-_{5,8}=\begin{bNiceMatrix}\text{N/A} \end{bNiceMatrix}, \quad
    M^-_{1,9}=\begin{bNiceMatrix} 1 & 0 \end{bNiceMatrix}, \quad
    M^-_{4,10} =\begin{bNiceMatrix} 1 & 0 \end{bNiceMatrix}, \\
    M^-_{3,9}=\begin{bNiceMatrix} 0 & 1 \\ 0 & 0 \end{bNiceMatrix}, \quad
    M^-_{6,10}=\begin{bNiceMatrix} 0 & 1 \\ 0 & 0 \end{bNiceMatrix}
\end{array}
\end{align*}

\begin{figure}[htbp]
    \centering
    \includegraphics[width=\linewidth]{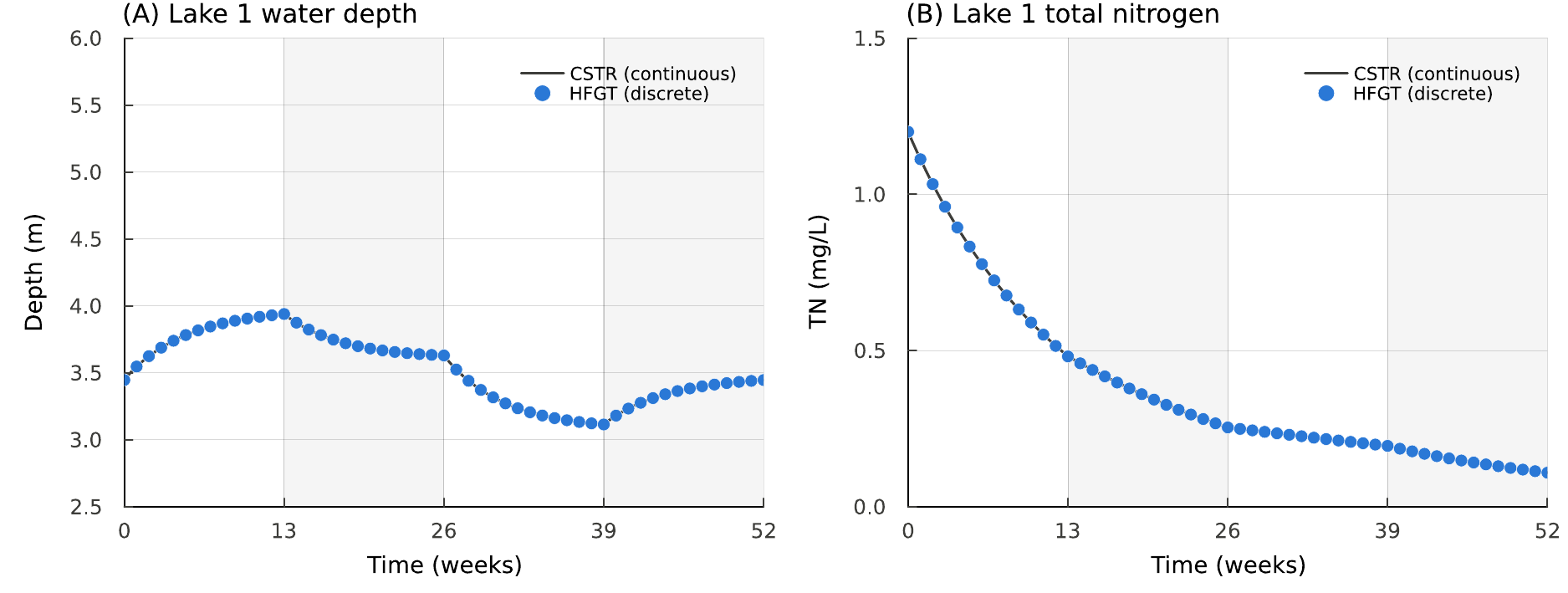}
    \caption{Simulation results for the system with one lake: \textbf{(A)} Lake 1 water depth and \textbf{(B)} Lake 1 total nitrogen, computed by the discrete-time HFGT formulation (markers, shown weekly) and the continuous-time CSTR equations (line). Shading marks the April--June and October--December input seasons.}
    \label{fig:sim1results}
\end{figure}

The simulation results for Illustrative Example 1 are presented in Fig.~\ref{fig:sim1results}, together with the continuous-time solution of the corresponding CSTR equations (Eqs.~\ref{Eq:dif-MB-water}--\ref{Eq:dif-mixing}, extended to the river reach and receiving water). Both solutions use identical parameters, read from the same instantiated model. Lake 1 rises during the wet winter and spring and falls during the dry summer, while its total nitrogen declines from 1.2 to 0.11~mg~L$^{-1}$ as low-nitrogen inflow flushes the lake; all of the net water input leaves the lake and reaches the outlet over the year. The discrete HFGT and continuous CSTR solutions agree to within 0.08~cm in depth and 0.21\% in total nitrogen (root-mean-square difference 0.00034~mg~L$^{-1}$). Because the comparison exercises the resistance-based outflow and the mixing constraint as well as the mass balance, this agreement verifies that the toolbox-generated incidence matrices, the device models, and the Julia implementation together reproduce the conventional formulation. This simple system serves as a baseline before progressing to spatially distributed configurations.

\subsection{Illustrative Example 2: A System with Three Lakes}

\begin{figure}[htbp]
    \centering
    \includegraphics[width=0.9\linewidth]{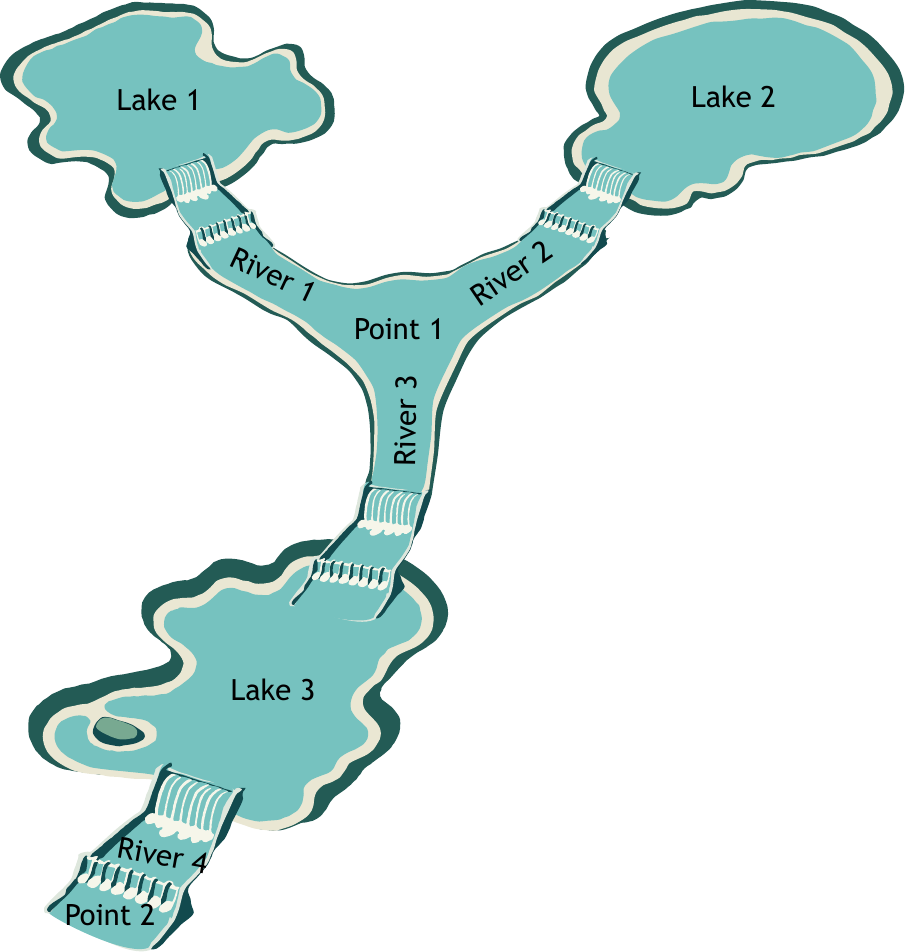}
    \caption{Diagram of the system with three lakes.}
    \label{fig:diagramThreeLakes}
\end{figure}

The three-lake system introduces additional complexity by connecting three lakes, each with its own inflow. Lakes 1 and 2 drain to Point 1, which drains to Lake 3, which in turn drains to Point 2 below the watershed outlet (Fig.~\ref{fig:diagramThreeLakes}). This example demonstrates the scalability of the MBSE--HFGT workflow and the need for computational modeling to derive and solve the associated equations.

\begin{figure}[htbp]
    \centering
    \includegraphics[width=\linewidth]{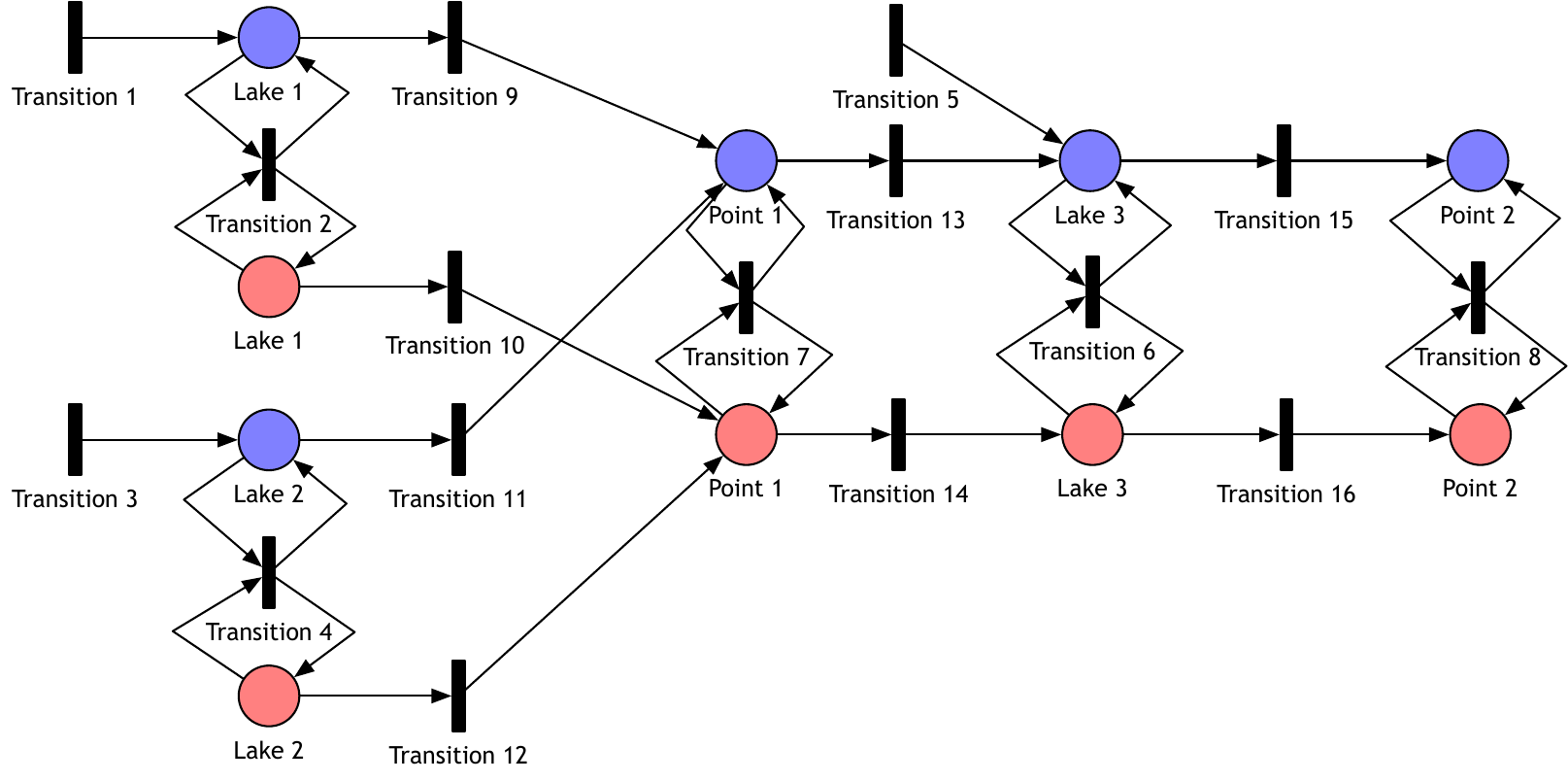}
    \caption{Elementary Petri net for the system with three lakes.}
    \label{fig:s2-elementary}
\end{figure}

The instantiated architecture for this configuration builds upon the watershed reference architecture and is visualized directly through the elementary Petri net in Fig.~\ref{fig:s2-elementary}, where each buffer--operand pair is represented by a distinct place. The XML file defining this system was processed using the HFGT toolbox, which produced the hetero-functional incidence tensors and system attributes used by the Julia-based simulator. The lakes differ in surface area, depth, and contributing area (Section~\ref{sec:parameters}), and therefore in their hydrological response.

\begin{figure}[htbp]
    \centering
    \includegraphics[width=\linewidth]{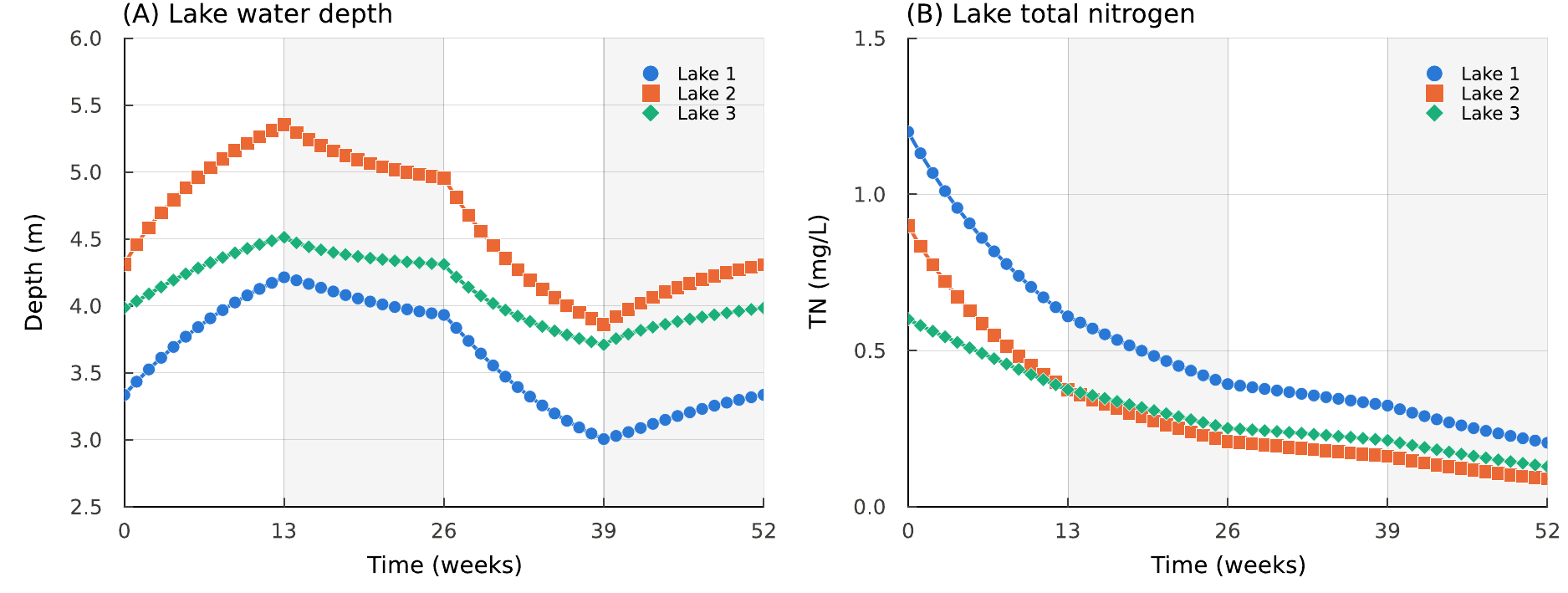}
    \caption{Simulation results for the system with three lakes: \textbf{(A)} lake water depth and \textbf{(B)} lake total nitrogen. Shading marks the April--June and October--December input seasons.}
    \label{fig:sim2results}
\end{figure}

The results are shown in Fig.~\ref{fig:sim2results}. Lake depths follow the seasonal input, varying over the year by 1.2, 1.5, and 0.8~m in Lakes 1, 2, and 3. Lake 2, with the largest contributing area, varies most, while Lake 3 receives the combined, routed outflow of the upstream lakes and responds more smoothly. As in Example 1, nitrogen is flushed from each lake by low-nitrogen inflow, falling to 0.09--0.20~mg~L$^{-1}$ by the end of the year, and all of the net water input reaches the outlet.

\subsection{Illustrative Example 3: A System with Three Lakes and Three Land Segments}

\begin{figure}[htbp]
    \centering
    \includegraphics[width=0.9\linewidth]{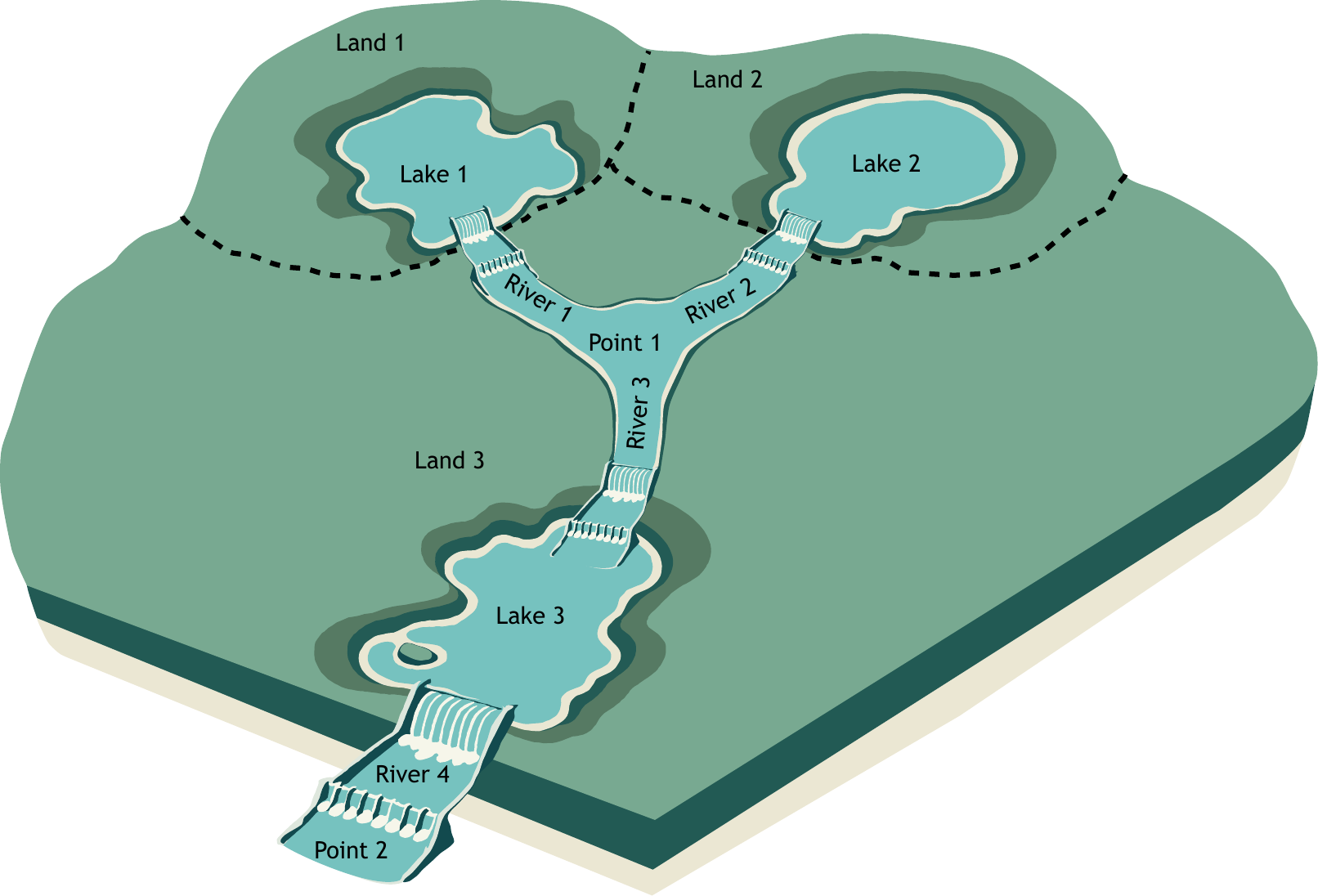}
    \caption{Diagram of the system with three lakes and three land segments.}
    \label{fig:diagramThreeLakesLand}
\end{figure}

This final example adds land segments to the three-lake system, introducing a new element of the reference architecture. Each land segment receives precipitation and nitrogen, stores both, and drains to its lake, coupling land and water systems (Fig.~\ref{fig:diagramThreeLakesLand}). The nitrogen applied to each land segment is an exogenous input here, but it represents a human decision, the application of fertilizer; in a coupled model it would be supplied by a model of agricultural production rather than specified directly. This configuration demonstrates the extensibility of the MBSE--HFGT workflow to coupled systems spanning multiple environmental domains.

\begin{figure}[htbp]
    \centering
    \includegraphics[width=\linewidth]{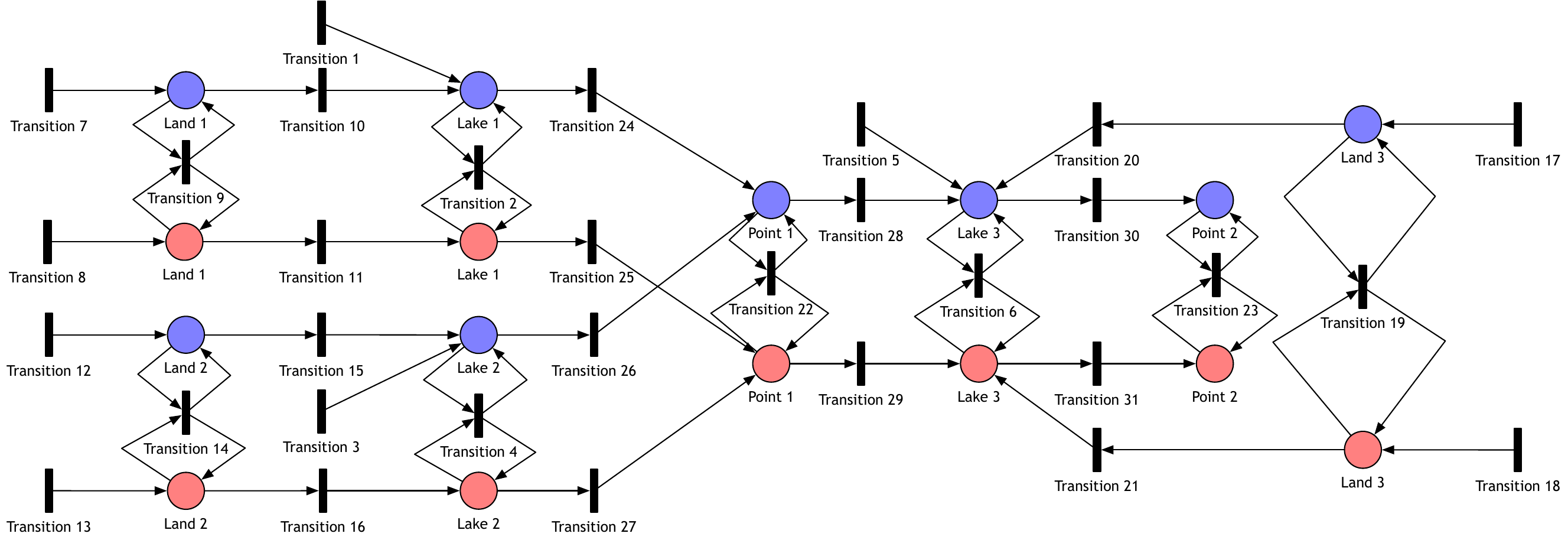}
    \caption{Elementary Petri net for the system with three lakes and three land segments.}
    \label{fig:s3-elementary}
\end{figure}

The instantiated architecture for this system is represented by the Petri net in Fig.~\ref{fig:s3-elementary}. The land segments introduce nitrogen-rich runoff, which is tracked alongside water flows. Each land segment has the area of its lake's contributing area in Example 2, so the two examples receive the same total water input, and the three segments receive nitrogen at the loading rates of soybeans (Land 1), other agronomic crops (Land 2), and pasture (Land 3) (Section~\ref{sec:parameters}).

\begin{figure}[htbp]
    \centering
    \includegraphics[width=\linewidth]{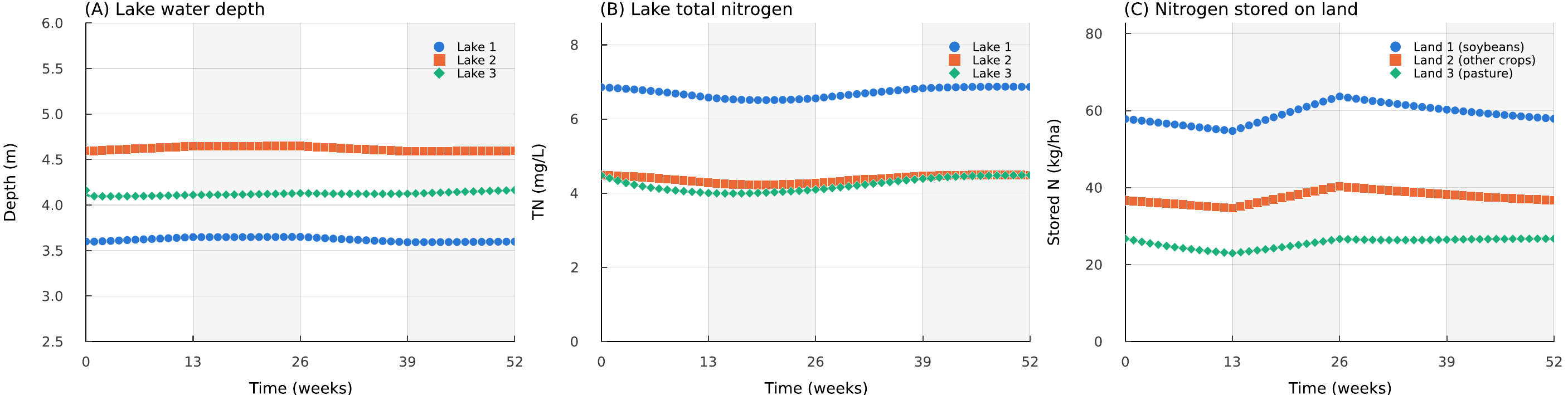}
    \caption{Simulation results for the system with three lakes and three land segments: \textbf{(A)} lake water depth, \textbf{(B)} lake total nitrogen, and \textbf{(C)} nitrogen stored on the land segments. Land 1, 2, and 3 drain to Lakes 1, 2, and 3 and receive nitrogen at the loading rates of soybeans, other agronomic crops, and pasture. Shading marks the April--June and October--December input seasons.}
    \label{fig:sim3results}
\end{figure}

Simulation results are presented in Fig.~\ref{fig:sim3results}. Nitrogen stored on the land segments rises during the spring application period and declines over the rest of the year, ranging from 55--64~kg~ha$^{-1}$ on Land 1 to 23--27~kg~ha$^{-1}$ on Land 3 in proportion to the loading rates. The lakes integrate these loads: total nitrogen remains between 4.0 and 6.9~mg~L$^{-1}$, highest in Lake 1, which drains the soybean segment, and varies seasonally by less than 0.5~mg~L$^{-1}$. Over the year, 44{,}510~kg of nitrogen enter the system, 44{,}404~kg leave through the outlet, and storage changes by 105~kg; this balance closes to machine precision (relative error $1.7\times10^{-16}$), confirming mass conservation in the instantiated formulation.

Representing the land segments explicitly also changes the lakes' hydrological response. Although Examples 2 and 3 receive the same total water input, lake depths in Example 3 vary by only 0.06--0.07~m over the year, compared with 0.8--1.5~m in Example 2. Water stored on the land segments drains to the lakes over weeks to months, so the seasonal pattern of the input reaches the lakes attenuated. Because each land segment and its lake are connected through a low fluidic resistance, their water levels move together and the lake outlet effectively drains their combined storage. This illustrates how adding an element of the reference architecture can alter the dynamics of the elements it connects to, without any change to the underlying formulation.

\section{Discussion} \label{sec:discussion}
The three examples presented throughout this paper demonstrate how the MBSE--HFGT methodology systematically transforms a conceptual system description into an executable mathematical model. Beginning with a single-lake example and progressing to increasingly complex watershed configurations, each example introduced additional system structure while preserving the same underlying modeling workflow. Rather than requiring new mathematical formulations for each application, the framework accommodates increasing system complexity through the addition of model elements within a consistent ontology and reference architecture.

A central feature of the methodology is the separation of system structure from system physics. The Engineering System Net captures the topology of the system and the conservation relationships among materials, while application-specific device models describe the physical behavior governing individual processes. This separation allows the same structural representation to support different physical models without modifying the underlying network formulation. As a result, discipline-specific process models can be incorporated, replaced, or refined independently while maintaining a consistent mathematical framework.

The examples also illustrate the scalability of the MBSE--HFGT approach. Expanding from a single lake to multiple lakes and agricultural land segments required only the addition of new model instances rather than changes to the modeling methodology itself. The HFGT toolbox automatically generated the corresponding tensors and optimization structures from the system description, demonstrating how increasingly complex environmental systems can be synthesized while preserving structural transparency.

Although the hydrological systems considered in this paper are intentionally simplified, they capture many of the defining characteristics of environmental and infrastructure systems, including material storage, transport, mixing, and transformation. Consequently, the modeling workflow presented here extends naturally beyond watershed applications to larger environmental systems and, more broadly, to other engineering domains. In particular, human systems can be coupled to the hydrological model through the capabilities they act on: the land segment's capability to accept nitrogen is where an agricultural or economic model would supply fertilizer inputs. Economic input--output models have already been represented in MBSE--HFGT with results identical to their traditional formulation~\cite{naderi:2026:00}, so coupling them to watershed models requires only connecting their sector outputs to these capabilities.

\subsection{Limitations, Pitfalls, and Troubleshooting} \label{sec:limitations}
The examples in this paper are intentionally simplified, and several assumptions limit their direct applicability. All parameters and inputs are synthetic, so the examples demonstrate that the HFNMCF formulation reproduces the conventional mass-balance model but do not validate either model against observations. Nitrogen is treated as conservative, with no representation of in-lake processes such as denitrification, settling, or biological uptake. Consequently, lake concentrations in Example 3 approach those of the land runoff. Flow follows a linear resistance law, and nitrogen is assumed to mix completely. The Eulerian control-volume formulation also omits operand nets, so processes that change an operand's internal state, such as nitrogen transformation, require additional modeling. A further limitation arises from the resistance law, which drives flow according to the difference in hydraulic head between connected storage elements. As a result, runoff from a land segment slows as the receiving lake rises. Because hillslope runoff is largely independent of lake level, the attenuation of lake-level variations in Example 3 should be interpreted with this limitation in mind.

Several practical pitfalls emerged during model instantiation. First, capability execution is assumed to be instantaneous (\(k_{d\psi}=0\)), and explicit time-stepping of the Engineering System Net is accurate only when \(\Delta T\) is sufficiently short relative to the fastest-responding storage element. In Example 3, a daily time step caused the river reach at Point 1 to overfill within a single step, switching off upstream lake outflows on alternate days. Reducing the time step to 6 hours eliminated this behavior; results at this resolution differed from those at much finer time steps by less than 0.2\% in nitrogen and 1~cm in depth. Comparing results across two time steps therefore provides a simple, effective numerical check. Second, when connected storage elements differ in elevation by more than their typical depths, the resistance law can produce nearly constant outflows that drain an element below its available storage. Elevations should therefore be chosen so that outflows remain governed by storage, as in the examples (Section~\ref{sec:parameters}). Third, expressing mixing constraints in terms of concentrations (\(m/V\)) can cause numerical instability as buffer volumes approach zero. The equivalent formulation \(m\dot{V}=V\dot{m}\) avoids division by volume (Section~\ref{sec:hfgtmath}). Finally, inconsistent initial storage states can produce transients that are mistaken for meaningful system behavior. Initializing the model from a spin-up using the same time step helps avoid this problem.

\subsection{Key Takeaways and Good Modeling Practices} \label{sec:keytakeaways}
\paragraph*{Model form and function before mathematics}
Rather than beginning with governing equations, the MBSE--HFGT workflow begins by explicitly modeling the system's form and function. The reference architecture captures the system's structural elements, the capabilities they perform, and the operands upon which they act. Once these relationships are established, the Engineering System Net and HFNMCF formulation can be derived systematically. This separation makes the resulting models easier to understand, extend, and maintain.

\paragraph*{Separate conservation laws from constitutive behavior}
One of the principal strengths of the HFNMCF formulation is its separation of conservation laws from constitutive behavior. The Engineering System Net enforces mass conservation and the structural relationships among system elements, while application-specific physics are incorporated through device model constraints. Maintaining this separation enables the same system architecture to support alternative hydrological, ecological, or other discipline-specific process models without altering the underlying mathematical framework.

\paragraph*{Develop models incrementally}
Complex engineering systems are most effectively developed through successive refinement. Beginning with a simple conceptual model allows the reference architecture and its corresponding mathematical formulation to be validated before additional resources, processes, and operands are introduced. The examples presented in this paper demonstrate how increasingly complex system instances can be constructed from a common reference architecture. As new application domains are considered, the reference architecture and ontology can be extended while preserving the underlying HFGT modeling workflow.

\paragraph*{Reuse reference architectures whenever possible}
Reference architectures provide reusable templates for recurring classes of systems. Once developed, they reduce modeling effort, encourage consistency across applications, and facilitate collaboration among multiple model developers. The watershed reference architecture developed in this paper demonstrates how domain knowledge can be encoded once and reused across multiple watershed configurations.

Collectively, these practices promote models that are more transparent and extensible as engineering systems increase in scale and complexity.

\section{Conclusion} \label{sec:conclusion}
This paper presented a systematic workflow for developing executable environmental system models using an integrated MBSE--HFGT methodology. Through an illustrative hydrologic system case study, the paper demonstrated how conceptual system descriptions can be translated into SysML-based reference architectures, Engineering System Nets, and ultimately the HFNMCF formulation. Although the governing equations remained consistent with conventional process-based modeling, the MBSE--HFGT framework made the underlying system structure explicit.

Beyond the illustrative watershed examples, the methodology is broadly applicable to heterogeneous engineering systems composed of interacting materials, processes, and infrastructure. Previous applications of HFGT have demonstrated its use in electric power systems, water distribution, wastewater treatment, transportation, manufacturing, healthcare, economic input--output systems, and other system-of-systems applications~\cite{Schoonenberg:2019:00, Schoonenberg:2022:00, Farid:2022:ISC-J49, Farid:2016:ISC-BC06, Thompson:2022:00, Khayal:2021:00, harris:2026:00,naderi:2026:00}. The methodology as presented here provides a practical entry point for applying these concepts within environmental engineering and establishes a repeatable workflow that can be adapted to many other application domains.

Although the examples presented here are intentionally simplified, the methodology has already been extended to a realistic, data-driven representation of the Chesapeake Bay Watershed~\cite{harris:2026:00} using the Chesapeake Assessment Scenario Tool (CAST)~\cite{Hood:2021:00}. This progression illustrates how the modeling workflow developed in this paper scales from pedagogical examples to operational environmental systems. Future work will continue to expand the HFGT toolbox, improve interoperability with domain-specific modeling software, and integrate additional ecological, economic, and social subsystems within a unified system-of-systems framework.

More broadly, this methods paper demonstrates that HFGT is not intended to replace traditional engineering models but rather to provide a structured framework for integrating them. By making system form and function explicit, the MBSE--HFGT methodology offers a transparent and extensible foundation for modeling the increasingly interconnected engineering systems that characterize the Anthropocene~\cite{Little:2023:ISC-J50,Little:2026:ISC-JR10}.


\section*{Conflict of Interest Statement}
The authors declare that the research was conducted in the absence of any commercial or financial relationships that could be construed as a potential conflict of interest.

\section*{Author Contributions}
MSH: Conceptualization, Methodology, Software, Validation, Formal analysis, Investigation, Writing -- original draft, Writing -- review \& editing, Visualization, Project administration.
MMN: Writing -- review \& editing.
EG: Writing -- review \& editing.
JCL: Conceptualization, Resources, Writing -- review \& editing, Supervision, Project administration, Funding acquisition.
AMF: Conceptualization, Methodology, Software, Validation, Writing -- review \& editing, Supervision, Project administration, Funding acquisition.

\section*{Funding}
This research was supported by the Growing Convergence Research Program of the National Science Foundation under Grant Numbers OIA 2317874 and OIA 2317877.

\section*{Acknowledgments}
This research is based on work supported by the Growing Convergence Research Program of the National Science Foundation under Grant Numbers OIA 2317874 and OIA 2317877. 

\section*{Generative AI Statement}
During the preparation of this work the authors used Claude (Anthropic) to assist with code development, minor figure revision, and manuscript editing. The authors reviewed and edited all code, figures, and text and take full responsibility for the content of the publication.

\section*{Data Availability Statement}
All software and data used to generate the illustrative simulation results are openly available at \url{https://github.com/LIINES/HFGT-Hydrology-Model}. 
The repository includes the HFGT toolbox input XML files, the version of the HFGT toolbox used to generate the system tensors, the output HDF5 files for each simulation, the Julia scripts used to instantiate, simulate and visualize the models, and the scripts used to generate every figure in this paper. The repository also records the exact Julia package versions and documents how to rerun every simulation and regenerate the toolbox outputs. All case-study data were synthetically generated for illustrative purposes. Net water inputs were scaled to the NOAA 1991--2020 climate normal precipitation for Blacksburg, Virginia (1083~mm~yr$^{-1}$)~\citep{NOAA:2021:00}, of which approximately 40\% was taken as water yield, the remainder being returned by evapotranspiration~\citep{Sanford:2013:00}. Nitrogen inputs to land segments used Chesapeake Bay Program Phase~6 Watershed Model edge-of-stream total-nitrogen loading rates for full-season soybeans, other agronomic crops, and pasture~\citep{CBP:2018:00}, and initial lake concentrations were chosen above the recommended lake total-nitrogen criterion for nutrient ecoregion IX~\citep{EPA:2000:lakesIX}. No proprietary, restricted, or observational datasets were used in this study.

\bibliographystyle{Frontiers-Harvard}
\bibliography{references}
\end{document}